\documentclass[fleqn,usenatbib]{mnras}

\usepackage{newtxtext,newtxmath}

\usepackage[T1]{fontenc}
\DeclareRobustCommand{\VAN}[3]{#2}
\let\VANthebibliography\thebibliography
\def\thebibliography{\DeclareRobustCommand{\VAN}[3]{##3}\VANthebibliography}

\usepackage{graphicx}	
\usepackage{amsmath}	
\usepackage{makecell}
\usepackage{subcaption}
\usepackage{natbib}
\usepackage{comment}

\graphicspath{{plots/}}  

\newcommand{\hMpc}{{\ifmmode{\,h^{-1}{\rm Mpc}}\else{$h^{-1}$Mpc}\fi}}
\newcommand{\hkpc}{{\ifmmode{\,h^{-1}{\rm kpc}}\else{$h^{-1}$kpc}\fi}}
\newcommand{\hMsun}{{\ifmmode{\,h^{-1}{\rm {M_{\odot}}}}\else{$h^{-1}{\rm{M_{\odot}}}$}\fi}}
\newcommand{\Msun}{\,\rm {M_{\odot}}}
\newcommand{\Mstar}{{\ifmmode{\,M_{*}}\else{$M_{*}$}\fi}}
\newcommand{\Mhalo}{{\ifmmode{\,M_{\rm halo}}\else{$M_{\rm halo}$}\fi}}
\newcommand{\ltsima}{$\; \buildrel < \over \sim \;$}
\newcommand{\gtsima}{$\; \buildrel > \over \sim \;$}
\newcommand{\lsim}{\lower.5ex\hbox{\ltsima}}
\newcommand{\gsim}{\lower.5ex\hbox{\gtsima}}

\newcommand{\theth}{\textsc{The Three Hundred}}

\newcommand{\gadgetx}{\textsc{Gadget-X}}
\newcommand{\gadget}{\textsc{Gadget3}}
\newcommand{\gadgetb}{\textsc{Gadget2}}
\newcommand{\simba}{\textsc{Gizmo-Simba}}

\def\kms{\,{\rm km}\,{\rm s}^{-1}}

\defcitealias{meneghetti2020excess}{M20}

\title[$M_{sub} - V_{circ}$ relation]{$\theth$: $M_{sub} - V_{circ}$ relation}

\author[Srivastava et al.]{
Atulit Srivastava,$^{1,2}$\thanks{E-mail: ATULIT.SRIVASTAV@ESTUDIANTE.UAM.ES}
Weiguang Cui,$^{1,2,3}$\thanks{E-mail: weiguang.cui@uam.es; Talento-CM fellow}
Massimo Meneghetti,$^{4,5}$
Romeel Dave, $^{3,5}$
Alexander Knebe, $^{1,2,6}$
\newauthor
Antonio Ragagnin,$^{7,8,9}$
Carlo Giocoli, $^{4,5}$
Francesco Calura,$^{4}$
Giulia Despali,$^{7,10}$
Lauro Moscardini$^{4,5,7}$
\newauthor
and Gustavo Yepes$^{1,2}$\\
$^{1}$Departamento de Física Teórica, M-8, Universidad Autónoma de Madrid, Cantoblanco 28049, Madrid, Spain\\
$^{2}$Centro de Investigación Avanzada en Física Fundamental (CIAFF), Universidad Aut\'{o}noma de Madrid, Cantoblanco, 28049 Madrid, Spain\\
$^{3}$Institute for Astronomy, University of Edinburgh, Royal Observatory, Edinburgh EH9 3HJ, United Kingdom\\
$^{4}$INAF-Osservatorio di Astrofisica e Scienza dello Spazio di Bologna, Via Piero Gobetti 93/3, 401 29 Bologna, Italy\\
$^{5}$INFN-Sezione di Bologna, Viale Berti Pichat 6/2, 401 27, Bologna, Itlay\\
$^{6}$ International Centre for Radio Astronomy
Research, University of Western Australia, 35 Stirling Highway, Crawley, Western Australia 6009, Australia\\
$^{7}$ Dipartimento di Fisica e Astronomia "Augusto Righi", Alma Mater Studiorum Università di Bologna, via Gobetti 93/2, I-40129 Bologna, Italy\\
$^{8}$ INAF-Osservatorio Astronomico di Trieste, via G. B. Tiepolo 11, I-34143 Trieste, Italy \\
$^{9}$ IFPU – Institute for Fundamental Physics of the Universe, Via Beirut 2, I-34014 Trieste, Italy \\
$^{10}$ Institut für Theoretische Astrophysik, Zentrum für Astronomie, Heidelberg Universität, Albert-Ueberle-Str. 2, 69120, Heidelberg, Germany
}

\date{Accepted XXX. Received YYY; in original form ZZZ}

\pubyear{2023}

\begin{document}
\label{firstpage}
\pagerange{\pageref{firstpage}--\pageref{lastpage}}
\maketitle

\begin{abstract}
In this study, we investigate a recent finding based on strong lensing observations, which suggests that the sub-halos observed in clusters exhibit greater compactness compared to those predicted by $\Lambda$CDM simulations.
To address this discrepancy, we performed a comparative analysis by comparing the cumulative mass function of sub-halos and the $M_{\text{sub}}$-$V_{\text{circ}}$ relation between observed clusters and 324 simulated clusters from $\theth$ project, focusing on re-simulations using $\gadgetx$ and $\simba$ baryonic models.
The sub-halos' cumulative mass function of the $\simba$ simulated clusters agrees with observations, while the $\gadgetx$ simulations exhibit discrepancies in the lower sub-halo mass range possibly due to its strong SuperNova feedback. Both $\gadgetx$ and $\simba$ simulations demonstrate a redshift evolution of the sub-halo mass function and the $V_{max}$ function, with slightly fewer sub-halos observed at lower redshifts. Neither the $\gadgetx$ nor $\simba$(albeit a little closer) simulated clusters' predictions for the $M_{\text{sub}}$-$V_{\text{circ}}$ relation align with the observational result. Further investigations on the correlation between sub-halo/halo properties and the discrepancy in the $M_{\text{sub}}$-$V_{\text{circ}}$ relation reveals that the sub-halo's half mass radius and galaxy stellar age, the baryon fraction and sub-halo distance from the cluster's centre, as well as the halo relaxation state play important roles on this relation.
Nevertheless, we think it is still challenging in accurately reproducing the observed $M_{\text{sub}}$-$V_{\text{circ}}$ relation in our current hydrodynamic cluster simulation under the standard $\Lambda$CDM cosmology. 
\end{abstract}




\section{Introduction}
Cold dark matter (CDM) plays an essential role in the formation and evolution of galaxies and galaxy clusters.
It can be detected solely through its gravitational effects, such as the bending of the light from background galaxies.
Galaxy clusters are gravitationally bounded systems with a mass around $10^{14}$ to $10^{15}$ solar masses, and dark matter makes up approximately 80 per cent of their mass.
Gravity drives the process of structure formation, with haloes assembling hierarchically over time. Galaxy cluster haloes, in particular, are among the late-forming structures \citep{white1991galaxy, tormen1998assembly,giocoli2007improved}.
Inside galaxy clusters, hundreds to thousands of sub-halos are resident in local minimum potential \citep{springel2001populating,giocoli2010substructure}. These inner structures are known as sub-halos. Investigating and understanding these sub-halos will help us to understand galaxy cluster formation in detail.

The paper from \cite{meneghetti2020excess} (hereafter \citetalias{meneghetti2020excess}) studied the gravitational lensing properties of both the cluster halos and sub-halos of the cluster samples observed in the Cluster Lensing and Supernova Survey with Hubble \citep[CLASH][]{2012ApJS..199...25P} and Hubble Frontier Fields  \citep{2017ApJ...837...97L} and compared them to hydro-simulated galaxy clusters. In their study, \citetalias{meneghetti2020excess} discovered that the Galaxy-Galaxy Strong Lensing (GGSL) probability from simulation, reconstructed means of lensing tool of \cite{bergamini2019enhanced}, is significantly lower compared to the observed clusters. This finding indicates that the observed clusters have much higher GGSL probability than those from hydrodynamic simulations under the $\Lambda$CDM cosmology. To support their argument, \citetalias{meneghetti2020excess} used the maximum circular velocities, $V_{\text{circ}}$\footnote{We will use $V_{\text{circ}}$ to denote the maximum circular velocities throughout this paper.}, of sub-halos within galaxy clusters as a metric to assess the degree of compactness, as it directly reflects the sub-halo potential for producing the strong lensing events. This $V_{\text{circ}}$ is associated with the 1D-velocity dispersion $\sigma_{\rm o}$ by $V_{\text{circ}} = \sqrt{2} \sigma_{\rm o}$, which is selected as one of the parameters in the lens modelling analysis of \citetalias{meneghetti2020excess}.
They noticed that the sub-halos in observed clusters have higher $V_{\text{circ}}$ values when compared to sub-halo samples from mass-matched clusters in the cosmological hydrodynamic simulations by \cite{planelles2014role}.
These findings suggest that the galaxies in these observed clusters are more efficient in lensing background sources and potentially more concentrated. The discrepancy between simulation and observation results may arise from limitations in the simulation's resolution or the presence of systematics. It has been suggested that the simulation output is sensitive to mass resolution and tidal disruption \citep{van2018disruption,green2021tidal}, which could potentially impact sub-halos properties. However, \cite{Meneghetti2022} (see also \citealt{ragagnin2022galaxies}) found that the resolution does not affect the GGSL probability, which, however, seems sensitive to the galaxy formation model implemented in the simulations. Nevertheless, it is still difficult to simultaneously reproduce galaxies' stellar mass function and internal structure. Another possible explanation is that this issue arises from an inaccurate understanding of the nature of dark matter within the $\Lambda$CDM paradigm, which may necessitate the exploration of alternative models such as self-interacting dark matter (SIDM) models \citep{yang2021self,bhattacharyya2022signatures} and cold and sterile neutrino (SN) dark matter models \citep{despali2020lensing}.

Using the simulated galaxy clusters from the Hydrangea/C-EAGLE cosmological hydrodynamic simulations, \cite{bahe2021strongly} did a similar comparison to the observed clusters,  as in \citetalias{meneghetti2020excess}.
Only one simulated cluster from Hydrangea at redshift $z=0.4$ matches closely with the mass range of the observed sample presented in \citetalias{meneghetti2020excess}, with mass of $M_{\text{200c}}$\footnote{
$M_{\text{200c}}$ represents the mass within a radius denoted as $R_{\text{200c}}$, measured from the center of a galaxy cluster's potential, where this radius signifies the region with an average density that is 200 times the critical density of the Universe.}$>5 \times 10^{14} h^{-1} M_{\odot}$.
They claimed that sub-halos in this highly resolved simulation match well with the observations (see also another study by \citealt{robertson2021galaxy} for resolution impact on simulation generated lensing signal).
In their study, \cite{bahe2021strongly} determined that $V_{\text{circ}}$ is higher by a factor of $2$ in Hydrangea and is consistent with respect to the observation trend.
This increase in the offset of the maximum circular velocity was attributed to the inclusion of baryons in the simulations. 
The comparison made by \cite{bahe2021strongly} between simulations with and without baryonic matter (i.e. dark matter only) revealed that sub-halos with a higher fraction of baryonic matter exhibited higher $V_{\text{circ}}$, implying that dense stellar cores capable of sustaining tidal stripping play a major role in explaining the observed high lensing signals (\citealt{armitage2019cluster}, also see \citealt{bahe2019disruption,joshi2019trajectories}). 
Additionally, \cite{bahe2021strongly} also checked the result from the Illustris-TNG300 simulation (\cite{marinacci2018first,naiman2018first,nelson2019illustristng,pillepich2018first,springel2018first}), and argued that both Illustris-TNG300 and Hydrangea simulations predicted high $V_{\text{circ}}$ values for massive sub-halos located in the vicinity of the cluster centre.
Thus, \cite{bahe2021strongly} concluded that there is no evidence of a significant disagreement between the observed sub-halos concentrations and predictions from the CDM model.

On the contrary, \cite{ragagnin2022galaxies} examined the effect of various numerical setups (such as resolution and softening length) and AGN feedback scheme on the interior structure of cluster sub-halos using six simulated zoomed-in regions of Dianoga, and they found contrasted results with respect to \cite{bahe2021strongly}.
Their results suggested that regardless of the numerical configuration used, the sub-halos of simulated clusters were unable to reproduce the observed $M_{\text{sub}}-V_{\text{circ}}$ ($M_{\text{sub}}$, sub-halo mass) relation from \cite{bergamini2019enhanced}. This failure to reproduce the scaling relation was particularly evident for sub-halo masses $M_{\text{sub}}<10^{11} h^{-1} M_{\odot}$, which corresponds to the mass range of interest for galaxy-galaxy strong lensing (GGSL) events. 
The simulated sub-halos exhibited $V_{\text{circ}}$ values are approximately $30\%$ smaller compared to the observed scaling relation presented by \cite{bergamini2019enhanced}. This was also observed for Hydrangea simulations discussed in \cite{bahe2021strongly}.
The scaling relationship between $M_\text{sub}$ (mass of sub-halos) and $V_{\text{circ}}$ (circular velocity), as derived from simulations, shows good agreement with observations in the high mass range ($M_{\text{sub}}>4 \times 10^{11} h^{-1} M_{\odot}$). However, concerns have been raised by \cite{ragagnin2022galaxies} regarding the simulations' tendency to produce high stellar masses for sub-halos within this mass range. This discrepancy in stellar mass could potentially be a key factor contributing to the observed agreement in the $M_\text{sub}-V_{\text{circ}}$ relation for the high mass range, and it may be associated also with the Hydrangea simulations examined by \cite{bahe2021strongly}. Although the simulations can reproduce the correct scaling relationship between $M_\text{sub}$ and $V_{\text{circ}}$ by adjusting the AGN feedback strength, the resulting galaxies exhibit unrealistic properties, such as having larger stellar masses compared to observed galaxies. As demonstrated by \cite{ragone2018bcg}, both Hydrangea and IllustrisTNG simulations show excessively large stellar masses in the brightest cluster galaxies. 
Therefore, it is important to emphasize that both \cite{Meneghetti2022} and \cite{ragagnin2022galaxies} clearly stated that simulations are unable to simultaneously reconcile with the observed $M_\text{sub}-V_{\text{circ}}$ relationship in the two sub-halo mass regimes. 

We would like to point out that all these previous studies are limited by the number of cluster samples which can not draw statistically solid conclusions as well as no correlation studies. In a recent letter, Meneghetti et al. (in prep.), performed a ray-tracing analysis of 324 galaxy clusters from the \theth\footnote{\url{https://www.the300-project.org}} and found that the \simba\ version run developed denser stellar cores and boosted the galaxy-galaxy strong lensing probability by a factor of $\sim 3$ than its \gadgetx\ counterparts. 
In this companion paper, we also use the simulated galaxy clusters from the $\theth$ project, as detailed in \cite{cui2018three,cui2022three}, to compare with the observed $M_{\text{sub}} - V_\text{{circ}}$ relation reported in \citetalias{meneghetti2020excess}. Although our simulated clusters have a slightly lower mass resolution than \cite{planelles2014role} and about 100 times lower than the Hydrangea simulated clusters, they have a significant advantage in terms of a large sample size, a relatively wide extensive mass range and, importantly, two different baryon models. 
Our sample includes approximately 10 times more simulated clusters than the Hydrangea sample used in \cite{bahe2021strongly}, and roughly 15 times more than the Dianoga simulation used in \cite{ragagnin2022galaxies}.
These advantages allow us to statistically investigate and understand the discrepancy. 

The paper is structured as follows: In Section \ref{sec:Section2}, we provide an introduction to the $\theth$ galaxy cluster simulation with the Amiga Halo Finder (AHF) halo catalogue which was used to identify host-halos and their corresponding sub-halos. We will also explain our methodology for selecting the samples of host halos and their sub-halos. 
In Section \ref{sec:Section3}, we compare the sub-halos mass distribution of \citetalias{meneghetti2020excess} to three reference clusters with the predictions from the simulation and examine how it evolves with redshift.
In Section \ref{Section:Vmax-function}, we present the cumulative sub-halo $V_{\text{circ}}$ function for the simulations.  
In Section \ref{sec:Section4}, we compare the observed $M_\text{sub}$ and $V_{\text{circ}}$ relation reported in \cite{bergamini2019enhanced} with the one generated from the data set of simulated clusters in the $\theth$ project. 
We also examine the influence of sub-halo and host-halo properties on the  $M_\text{sub}-V_{\text{circ}}$ relationship. Finally, in Section \ref{sec:Section5} we will summarise our results.

\section{Simulations} \label{sec:Section2}
The $\theth$ project, introduced in \cite{cui2018three}, consists of an ensemble of 324 galaxy clusters that were modelled based on the extraction of a mass-complete sample with largest virial halo mass $M_{\text{vir}}\gtrsim 8 \times 10^{14} h^{-1} M_{\odot}$ at $z=0$ from the Dark Matter-only Multidark simulation \citep[MDPL2,][]{klypin2016multidark}.
The MDPL2 simulation employs periodic boundary conditions with a cubic side of $1\text{Gpc}/h$ and contains $3840^{3}$ dark matter particles, with a mass of $1.5\times 10^{9}$ $h^{-1} M_{\odot} $. This dark matter-only simulation adopts cosmological parameters $(\Omega_{M} = 0.307, \Omega_{B} = 0.048, \Omega_{\Lambda} = 0.693, h = 0.678, \sigma_8 = 0.823, n_{s} = 0.96
)$ based on the Planck observations from \cite{refId0}. Each selected cluster is placed at the centre of the re-simulated box inside a high-resolution spherical region with a radius of $15 h^{-1}\text{Mpc} $.
The regions are filled with gas and dark matter particles (with $m_{\text{DM}} = 1.27 \times 10^{9} h^{-1} M_{\odot}$ and $m_{\text{gas}} = 2.36 \times 10^{8} h^{-1} M_{\odot}$) based on the original dark matter distribution, in accordance to the cosmological baryon fraction $\Omega_{B} = 0.048$. 
Beyond the $15 \hMpc$ range, the outer region is populated with low-resolution mass particles to simulate any large-scale tidal effects similar in a computationally efficient way compared to the original MDPL2 simulation. Subsequently, the 324 selected regions undergo re-simulation using different baryonic models and codes, namely $\gadgetx$ \citep{rasia2015cool} and $\simba$ \citep{dave2019simba, cui2022three}. For each simulated cluster in the $\theth$ project using $\gadgetx$ and $\simba$, we have 128 snapshot files corresponding to redshifts ranging from $z=17$ to $0$.

The details regarding the $\gadgetx$ and $\simba$ codes used for the re-simulation of clusters are as follows:
\begin{itemize}
    \item \textbf{$\gadgetx$}: It is an updated, modified version of $\gadget$ code \citep{murante2010subresolution,rasia2015cool,planelles2017pressure} in which the evolution of dark matter is followed by the gravity solver of the $\gadget$ Tree-PM code, an updated version of $\gadgetb$ code \citep{springel2005cosmological}. It incorporates an improved SPH scheme \citep{beck2016improved} with artificial thermal diffusion, time-dependent artificial viscosity, high-order Wendland C4 interpolating kernel, and wake-up scheme.
    The technique described in \cite{wiersma2009effect} is used to compute gas cooling for an optically thin gas with consideration of the contribution of metals. Additionally, a uniform ultraviolet (UV) background is incorporated by adopting the approach outlined in \cite{haardt1995radiative}.
    Star formation in this work follows the approach described in \cite{tornatore2007chemical} and adopts the star formation algorithm presented by \cite{springel2003cosmological}. 
    This algorithm treats gas particles as multiphase, contributing to a self-regulating interstellar medium when their densities rise over a particular threshold. The star formation rate is determined solely by the gas density in this model. Stellar feedback, specifically supernova feedback, is implemented as a kinetic energy-driven scheme, following the prescription in \cite{springel2003cosmological}.
    Each star particle is treated as a single stellar population (SSP), and the evolution of each SSP is modelled following \cite{chabrier2003galactic} stellar evolution prescriptions. Metals from Type Ia and Type II supernovae, as well as from asymptotic giant branch phases, are taken into account in the simulation, with the code following the evolution of 16 chemical species.
    The growth of black holes and the implementation of AGN feedback in \gadgetx\ are based on the refined model presented in \cite{steinborn2015refined}. In this model, super-massive black holes grow via Eddington-limited Bondi-Hoyle-like gas accretion, with a distinction made between hot and cold components.
    \item \textbf{$\simba$} It is based on the GIZMO cosmological hydro-dynamical code \citep{hopkins2015new} with its mesh-less finite mass scheme and utilises the galaxy formation input physics of the state-of-the-art Simba simulation \citep{dave2019simba}. The baryon model was re-calibrated because \theth\ initial conditions have a lower resolution than the original SIMBA simulation, and both simulations had different objectives (cosmological run for SIMBA and galaxy cluster for $\theth$). The GRACKLE-3.1 library \citep{smith2017grackle} is utilised to implement the processes of radiative cooling, photon heating, and gas ionization. The spatially-uniform ultraviolet background model \citep{haardt2012radiative} and the self-shielding prescription, based on the approach by \cite{rahmati2013evolution}, are employed. Additionally, an $H_{2}$-based star formation model from the \textsc{Mufasa} \citep{dave2016mufasa} is included.     
    The star formation-driven galactic winds are implemented based on a decoupled two-phase model. This model is also based on \textsc{Mufasa}, but with an additional mass loading factor derived from \cite{angles2017cosmic}. The chemical enrichment model tracks eleven elements with metals enriched from supernovae Type Ia and Type II and asymptotic giant branch stars. The black hole accretion description is based on two models: torque limited accretion model for cold gas \citep{angles2015torque,angles2017gravitational} and hot gas accretion model based on \cite{bondi1952spherically}. It incorporates three AGN feedback modes: a kinetic subgrid model for both `radiative mode' and `jet mode' with bi-polar ejections, and a kinetic X-ray feedback model following \cite{choi2012radiative}. A more extensive discussion about the baryon model can be found in \cite{cui2022three,dave2016mufasa,dave2019simba}.
\end{itemize}
Apart from the disparities in their models, it is essential to recognise that the two codes differ in their objectives when comparing simulation outputs to observations. 
The $\gadgetx$ simulation is tuned to accurately reproduce the gas properties and relations observed in observations, such as the temperature–mass ($T-M$) and integrated Sunyaev-Zeldovich decrement vs. mass ($Y-M$) relations \citep[e.g.,][]{Li2020, Sayers2023, Li2023}. On the other hand, the $\simba$ simulation is calibrated to reproduce galaxy stellar properties, including the total stellar fraction, satellite stellar mass function, and Brightest Cluster Group (BCG) halo mass functions \citep[see][for comparisons between the two simulations]{Zhang2022,Cui2022EPJWC.25700011C,Ferragamo2023}.
Since the introduction of $\theth$ project in \cite{cui2018three}, several studies have used this data on many different projects, such as, \cite{haggar2020thethreehundred,ansarifard2020three, deAndres2022}. We refer the readers to these papers for more details about the project.
\subsection{The Halo and Sub-halo Catalogues}\label{sec:Section2.1}
The simulation data is analyzed using the AHF (Amiga Halo Finder) open-source software \citep{knollmann2009ahf} to generate halo/sub-halo catalogues. AHF identifies structures hierarchically within cosmological simulations. It detects and locates spherical over-density peaks in the density field of the simulation, consistently considering dark matter, stars, and gas particles. The physical properties of all identified halos are determined based on the gravitationally bound particles. Halo positions are determined based on the over-density peak and the radius $R_{200c}$. Additionally, sub-structures, referred to as sub-halos, are identified using the same process. Sub-halos are smaller gravitationally bound entities located within the radius $R_{200c}$ of a larger central structure termed the host halo.

AHF searches for connected overdensity regions within the radius $R_{200c}$ of the main halo. These regions are considered potential sub-halos. For each potential sub-halo, AHF determines whether the particles within the overdensity region are gravitationally bound to the main halo. This involves analyzing and comparing the particles' velocities with the local escape velocity obtained using the spherical potential approximation. If the overdensity region is found to be gravitationally bound to the main halo, it is confirmed as a sub-halo.
In the following subsection, we will describe the selection procedure of our host halos and their associated sub-halos used in our study. 

\subsection{Host-halo and Sub-halo Sample Selection}\label{sec:Section2.2}

\begin{table*}
	\centering
	\caption{Host halos and sub-halos samples obtained from the $\gadgetx$ simulated clusters dataset. The meaning of each column is indicated in the header (See Section \ref{sec:Section2.2} for further details). The information presented in the table pertains to sub-halos that are located at distances less than $0.15 R_{200c}$. Sub-halo mass threshold of $M_{\text{sub}}>1.27 \times 10^{11} h^{-1} M_{\odot}$ is used in the table to calculate the statistics.}
	\label{tab:table1}
	\begin{tabular}{ccccccccc} 
		\hline
		\thead{$z$} & \thead{$N_{\text{host}}$} & \thead{Median $M_{200c}$ \\ $[h^{-1} M_{\odot}]$} & \thead{Median \\$N^{\text{sub}}_{2D}$} &\thead{Median \\$M_{\text{sub}}^{2D}$$[h^{-1} M_{\odot}]$}
  & \thead{Median \\$N^{\text{sub}}_{3D}$} &\thead{Median \\$M_{\text{sub}}^{3D}$$[h^{-1} M_{\odot}]$} & \thead{Total \\$N^{\text{sub}}_{2D}$} & \thead{Total \\$N^{\text{sub}}_{3D}$} \\
		\hline
		0.394 & 90 & $7.97 \times 10^{14}$ & 10 & $2.62 \times 10^{11}$ & 3& $2.78 \times 10^{11}$ & 895 & 310\\
		0.194 & 180 & $8.17 \times 10^{14}$ & 9 & $2.70 \times 10^{11}$ & 3& $2.84 \times 10^{11}$ & 1719 & 576\\
		0 & 321 & $8.46 \times 10^{14}$ & 7 & $2.57\times10^{11}$ & 2&$2.49\times10^{11}$ & 2631 & 875\\
		\hline
	\end{tabular}
\end{table*}
\begin{table*}
	\centering
	\caption{Similar to \autoref{tab:table1}, but for \simba.}
	\label{tab:table2}
	\begin{tabular}{ccccccccc} 
		\hline
		\thead{$z$} & \thead{$N_{\text{host}}$} & \thead{Median $M_{200c}$ \\ $[h^{-1} M_{\odot}]$} & \thead{Median \\$N^{\text{sub}}_{2D}$}&\thead{Median \\$M_{\text{sub}}^{2D}$$[h^{-1} M_{\odot}]$} & \thead{Median \\$N^{\text{sub}}_{3D}$} &\thead{Median \\$M_{\text{sub}}^{3D}$$[h^{-1} M_{\odot}]$} & \thead{Total \\$N^{\text{sub}}_{2D}$} & \thead{Total \\$N^{\text{sub}}_{3D}$} \\
		\hline
		0.394 & 82 & $8.04 \times 10^{14}$ & 15& $2.42\times10^{11}$ & 6&$2.46\times10^{11}$ & 1264 & 488\\
		0.194 & 169 & $8.14 \times 10^{14}$ & 14& $2.35\times10^{11}$ & 5& $2.33\times10^{11}$& 2373 & 966\\
		0 & 302 & $8.32 \times 10^{14}$ & 12 &$2.30\times10^{11}$ & 5& $2.21\times10^{11}$& 3922 & 1578\\
		\hline
	\end{tabular}
\end{table*}

We selected the sample from each simulated cluster region (for both $\gadgetx$ and $\simba$ in $\theth$ dataset) focusing on three particular redshifts: $z=0.394$, $z=0.194$, and $z=0$. The redshift $z=0.394$ is primarily selected to enable a close comparison to observed galaxy clusters in \citetalias{meneghetti2020excess}, which have redshifts in the range $0.2<z<0.6$ with the median $z=0.39$. The two additional redshifts are for the purpose of evolution studies.
Host halos with $M_{\text{200c}}> 6.5\times 10^{14} h^{-1} M_{\odot}$ are selected in each simulation region, ensuring that the uncontaminated mass fraction of the high-resolution particles is greater than 0.98\footnote{The fraction is not 100 per cent, for AHF takes BH particles as low-resolution particles. However, changing this fraction does not affect our results.}.
This mass cut is to cover the observed cluster mass range, note that the three cluster masses in \citetalias{meneghetti2020excess} are: 1.59 $\pm$ 0.36 (MACS J1206.2–0847), 1.04 $\pm$ 0.22 (MACS J0416.1–0403) and 2.03 $\pm$ 0.67 (Abell S1063) $\times 10^{15} \Msun$ \citep[see Table 1 in][]{bergamini2019enhanced}.

For each host halo identified at three different redshifts in $\theth$ project's simulation runs, we further made selections of sub-halos with two scenarios given below:

\begin{itemize}
    \item [\textbf{2D}] Sub-halos that have $M_{\text{sub}}>1\times 10^{10} h^{-1} M_{\odot}$ and are located within a projected distance of less than $0.15 R_{\text{200c}}$ (where $R_{\text{200c}}$ represents the radius of the host halo) from the host-halo centre in the simulation's $XY$ plane, i.e., $R_{\text{2D}}<0.15 R_{\text{200c}}$.
    \item [\textbf{3D}] Sub-halos that have $M_{\text{sub}}>1\times 10^{10} h^{-1} M_{\odot}$ and are physically located at a distance less than $0.15 R_{\text{200c}}$ from their host-halo centre, i.e., $R_{\text{3D}}<0.15 R_{\text{200c}}$.
\end{itemize}
For the investigation of the cumulative sub-halo mass function and the $M_{\text{sub}}-V_{\text{circ}}$ relation, we considered the sub-halo mass cut mentioned above.
However, it's important to note that, in the correlation studies, we applied a significantly higher sub-halo mass cut of $M_{\text{sub}}>1.27\times 10^{11} h^{-1} M_{\odot}$ to the dataset. This was done to mitigate potential resolution-related issues that could affect the sub-halos properties.
Similarly, we also eliminated any contaminated sub-halos with a low-resolution particle mass fraction greater than 2 per cent. sub-halos without any stars are also excluded from our analysis.
Moreover, in \gadgetx, we found some sub-halos with unusually high stellar mass fraction ( $\it f_*$, is approximately above 0.8) but very low dark matter content close to the host halo's centre ($\approx R_{3D}<0.05 R_{200c}$), and subsequently excluded them from our analysis.
However, this issue was not observed in the sub-halos sampled from the $\simba$ runs.

The general information regarding our chosen sample is presented in Tables \ref{tab:table1} for \gadgetx\ and \ref{tab:table2} for \simba.
Table \ref{tab:table1} presents information on the selected host halos with the higher mass cut, including the number ($N_{\rm host}$, column 2), the median mass $M_{200c}$ (column 3) of host halos, the median number of sub-halos within $R_{2D}<0.15 R_{200c}$ of each host halo (column 4) and $R_{3D}<0.15 R_{200c}$ (column 6) with their median masses in column 5 and 6 respectively. 
Additionally, the table also provides the total number of selected sub-halos for $R_{2D}<0.15 R_{200c}$ (column 7) and $R_{3D}<0.15 R_{200c}$ (column 8) for $\gadgetx$ simulated clusters. The different rows show these quantities at the three different redshifts.
Similarly, Table \ref{tab:table2} reports information for the selected host-halos and sub-halos for simulated clusters at three different redshifts for the $\simba$ run. 

Based on this dataset of simulated clusters' host-halos and sub-halos from $\theth$ dataset, we will commence our investigation to examine whether significant offsets exist between the observations of \citetalias{meneghetti2020excess} and the simulations in the context of strong gravitational lensing. 

\section{sub-halo mass function} \label{sec:Section3}
\begin{figure*}
  \centering
  \includegraphics[width=\textwidth]{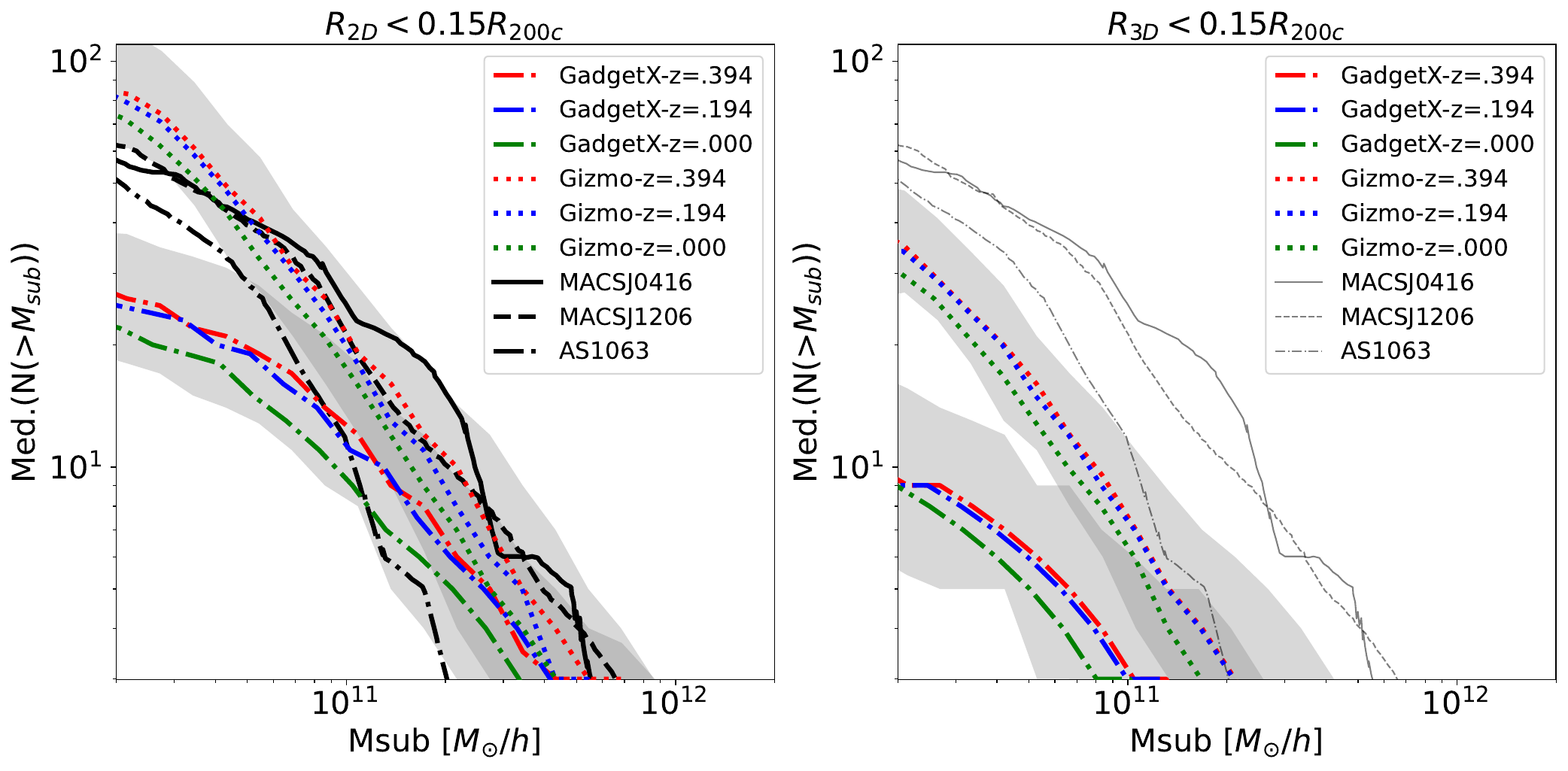}
  \caption{2D colored (left panel) and 3D (right panel) cumulative sub-halo mass functions. The dotted line style represents the $\simba$ simulation results, while dash-dot lines show median cumulative sub-halo mass functions from \gadgetx. The shaded areas show the $16^{th}-84^{th}$ percentiles from all clusters at $z=0.394$. The mass functions of sub-halos in $\gadgetx$ and $\simba$ simulations are displayed for three redshifts: $z=0.394$ (red), $z=0.194$ (blue), and $z=0$ (green). The projected results in the left panel used all sub-halos located within a projected 2D distance of $0.15 R_{\text{200c}}$, i.e., $R_{\text{2D}}<0.15 R_{\text{200c}}$. On the other hand, the right panel illustrates the results using only sub-halos situated within a physical 3D distance of $R_{\text{200c}}$, i.e., $R_{\text{3D}}<0.15 R_{\text{200c}}$. In both panels, the same observed sub-halo mass functions from three reference clusters in \protect\citetalias{meneghetti2020excess} are presented with black curves with different line styles; see the legend for details.}
  \label{Figure:1}
\end{figure*}

We begin our analysis by comparing the cumulative sub-halo mass functions predicted by $\theth$ clusters to the ones derived from the lens model of the three reference clusters, MACSJ0416, MACSJ1206, and AS1063, in \citetalias{meneghetti2020excess}.
We calculate the sub-halo mass function for each cluster to determine the median cumulative sub-halo mass function at the specified redshifts for the $\gadgetx$ and $\simba$ simulations. This is accomplished by utilising the available sub-halo information associated with each cluster. Next, we bin the sub-halos based on their mass, $M_{\text{sub}}$, into logarithmic mass bins and calculate the median value of $N(>M_{\text{sub}})$ for each bin.
This process yields the median cumulative sub-halo mass function for the simulated clusters at the respective redshifts.
Additionally, we calculate the lower and upper $34\%$ percentiles for $N(>M_{\text{sub}})$ in each logarithmic mass bin as their associated certainty.

Fig. \ref{Figure:1} depicts the median cumulative sub-halos mass function $R_{\text{2D}}<0.15 R_{\text{200c}}$ (left) and $R_{\text{3D}}<0.15 R_{\text{200c}}$ (right) for both $\gadgetx$ and $\simba$, at three redshifts, $z=0.394$, $z=0.194$, and $z=0$.
The grey shaded region in Fig. \ref{Figure:1} (left and right) represents the upper and lower $34\%$ quantiles for the cluster at redshift $z=0.394$ for $\gadgetx$ and $\simba$.
Despite employing different approaches, such as using projected 2D distance or considering the actual physical 3D distance between sub-halos and the host-halo centre, the cumulative sub-halo mass function follows a power-law trend when fitted analytically with a power law function, as previously demonstrated in \cite{giocoli2008population}.
The cumulative sub-halo mass function observed in the \simba\ simulation displays a clearly evident straight power law trend, with a power index almost equal to 1 when compared to the \gadgetx\ simulation.
Upon comparing the results with the observed sub-halo mass functions from \citetalias{meneghetti2020excess} obtained through a strong lensing model (represented by black curves with different line styles in \autoref{Figure:1}), we observe consistency between the observations of MACSJ0416 and MACSJ1206 and the results from $\simba$ simulated clusters within $R_{\text{2D}}<R_{\text{200c}}$.
For $\gadgetx$ simulated clusters, we find that the sub-halo mass function ($R_{\text{2D}}<0.15R_{200c}$) have a better agreement with observation results for sub-halo masses greater than $\sim 8 \times 10^{10} h^{-1} M_{\odot}$. Regarding the low sub-halo mass function at the low-mass end, its baryon model has a stronger resolution dependence\footnote{Though we show the sub-halo mass function down to $\sim 10^{10}h^{-1} M_{\odot}$, it is worth noting that these sub-halos only have around 10 dark matter particles.} because its sub-halo mass function is closer to the power-law if we don't apply the stellar mass constraint $M_{*}>0$  \citep[see also][which found many dark sub-halos in \gadgetx.]{Contreras2023}. 
In the scenario where sub-halos are situated within a radial distance of $R_{\text{3D}}<0.15R_{\text{200c}}$, both the $\gadgetx$ and $\simba$ simulations exhibit a lower median cumulative sub-halo mass function compared to the observed results. This discrepancy arises because the observational data inherently captures a 2D projection of the sub-halo distribution, while the condition $R_{\text{3D}}<0.15R_{\text{200c}}$ in the simulations takes into account the complete 3D spatial distribution of sub-halos. 
The projection effect increases the sub-halo numbers by a factor of $\sim 2.5$, regardless of the sub-halo masses. Note that we only considered sub-halos within $R_{200c}$ of the host halo for projection.
We verified that by applying this constraint (i.e., $R_{\text{3D}} < R_{200c}$), we only underestimate the sub-halo mass function for the case of $R_{\text{2D}} < 0.15 R_{\text{vir}}$ by approximately $2.21\%$ compared to the much larger radial constraint of $R_{\text{3D}} < 2.5 R_{200c}$.
Although the whole volume for the projection case is about five times larger than the 3D case, there are much fewer galaxies/sub-halos at large radius \citep[see][for example]{Li2020, Li2023}. Therefore, using a slightly larger projection distance will not affect this result much.
To perform this volume comparison, we directly compared the volume of a sphere having a radius of 0.15 $R_{\text{200c}}$ with the volume of a cylinder characterized by a radius of 0.15 $R_{\text{200c}}$ and a height of $R_{\text{200c}}$.
Lastly, there is a weak redshift evolution of the sub-halo mass functions in all the simulation samples \citep[see also][]{giocoli2008population,giocoli2010substructure}, which we will detail in the following subsection.

\subsection{The redshift evolution of the sub-halo mass function}
\begin{figure*}
\begin{subfigure}{\textwidth}
    \centering
    \includegraphics[width=\textwidth]{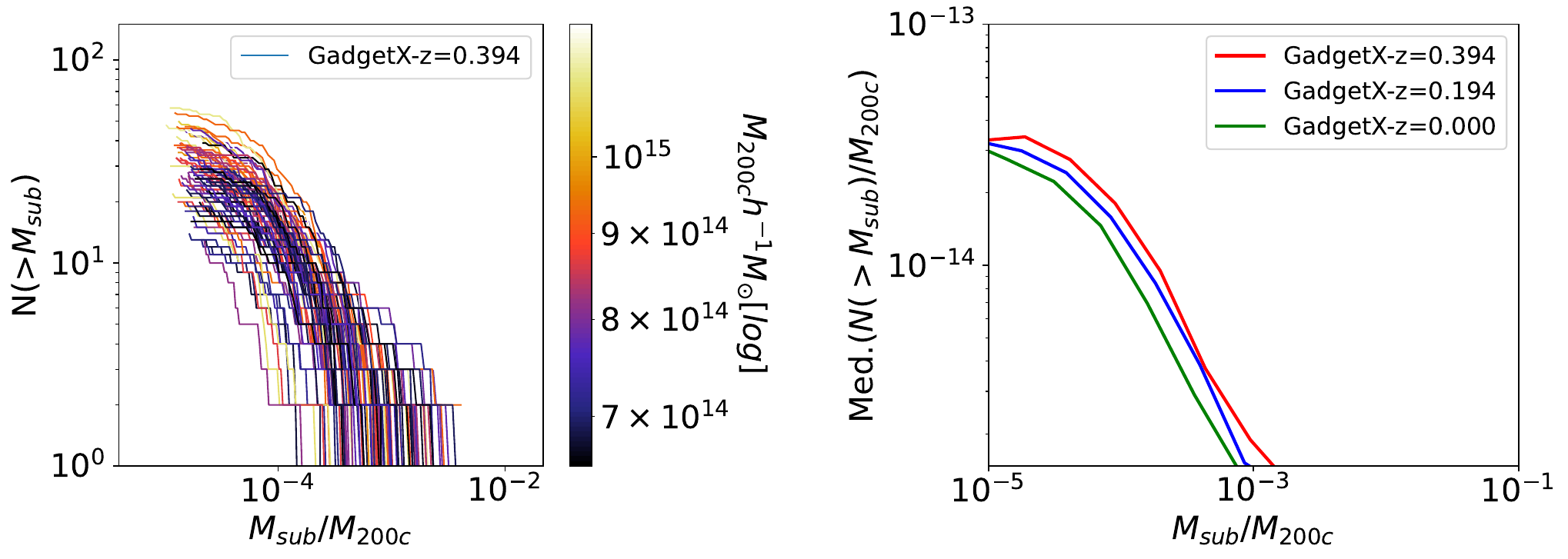}
    \caption{Left: the halo mass dependence of the sub-halo mass functions. Right: the redshift evolution of the cumulative sub-halos mass function after the normalisation of the halo mass. This is for the $\theth$ dataset from $\gadgetx$ simulated clusters.}
\end{subfigure}
\begin{subfigure}{\textwidth}
    \centering
    \includegraphics[width=\textwidth]{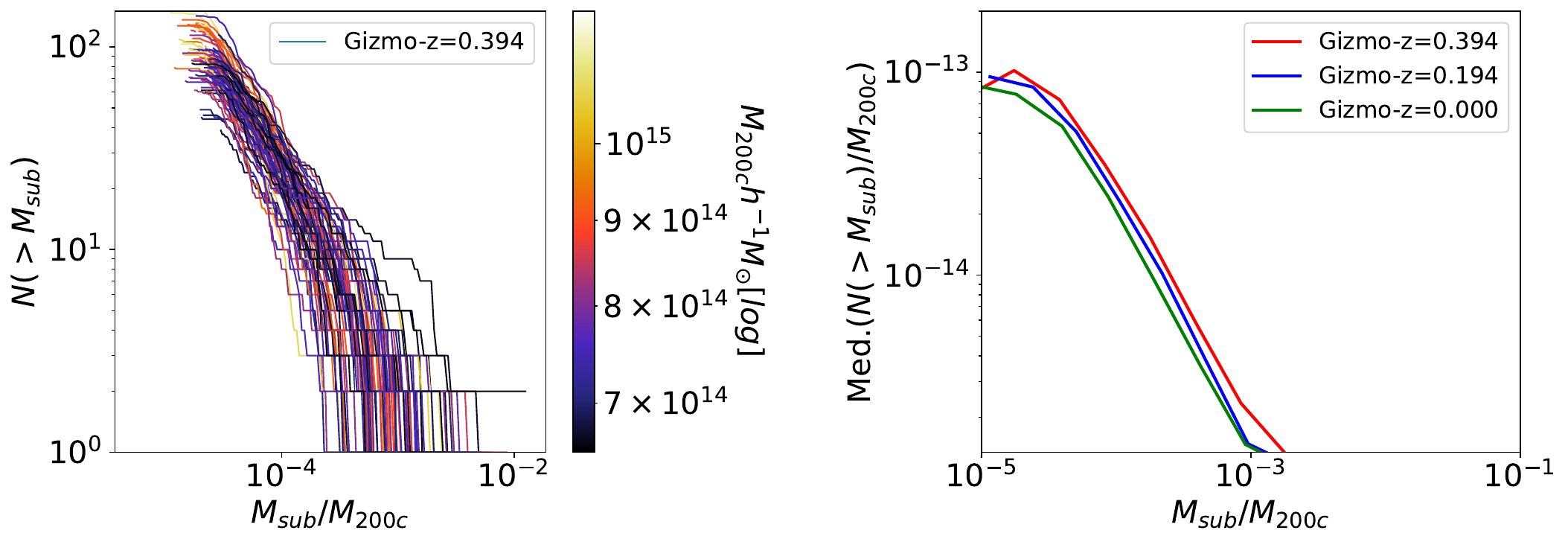}
    \caption{The same as the upper panel (a) but for \simba\ results.}
\end{subfigure}        
\caption{
In panels (a and b, left), we illustrate the unnormalized sub-halo mass functions at $z=0.394,$ demonstrating their dependence on halo mass. The right panels (a and b) display the corresponding cumulative sub-halo mass functions for the normalized sub-halo mass function at redshift $z=0.394$, along with the normalized sub-halo mass functions at $z=0.194$ and $z=0$. Further details regarding these plots are provided in sub-captions (a) and (b). 
}
\label{fig:figure 2}
\end{figure*}

It is expected that massive clusters have more sub-halos. As shown in \autoref{tab:table1} and \autoref{tab:table2}, the median halo mass slightly increases as redshift drops for both \gadgetx\ and \simba. One should expect a higher sub-halo mass function at $z=0$ than $z=0.394$. However, the results in \autoref{Figure:1} for both \gadgetx\ and \simba\ and within both $R_{2D}$ and $R_{3D}$ show an opposite evolution, i.e., a lower (fewer sub-halos) sub-halo mass function at $z=0$ compared to $z=0.394$. We suspect this could be due to the different halo mass distributions between these redshifts. Therefore, we investigate more on this in this subsection.

To examine the redshift evolution of the sub-halos mass function, we only present the projected results in the 2D case, which include more sub-halos. However, we note here that the $R_{3D}$ results are similar to the $R_{2D}$ ones.  
The sub-halos mass distribution for all $\gadgetx$ and $\simba$ simulated clusters at $z=0.349$ is shown in the left panels (both a and b) of Fig. \ref{fig:figure 2} with the line colour coding to the cluster mass, as indicated by the colour bar to its right. 
To show the residual redshift evolution of the sub-halo mass function, we first performed a normalisation step by dividing each host halo's cumulative sub-halo mass function by its own halo mass. This normalisation step eliminates any host halo mass dependence from the cumulative sub-halo mass distribution.
We then proceed with the calculation of the median sub-halos mass function by grouping the normalised sub-halos mass in logarithmic mass bins and then calculating median $N(>M_{sub})/M_{H}$ in each bin.
The right panel of Fig. \ref{fig:figure 2} shows the redshift evolution of the normalised sub-halo mass distribution predicted by the simulations for both $\gadgetx$ and $\simba$. 
The plot clearly illustrates the evolution of the sub-halo mass function as the redshift decreases from $z=0.349$ to $z=0$. The figure shows that, within a given parent halo, a greater number of sub-halos are expected to be observed at earlier times when they are dynamically young and less concentrated.
This similar inference about the redshift evolution of the sub-halos mass aligns with the findings presented in \cite{gao2004subhalo} and \cite{gao2011statistics}.
However, here we notice that the evolution we observed in our hydro-dynamical simulations is milder compared to the earlier studies of dark matter-only simulations.
Interestingly, the same evolution trend for hydro-dynamical simulations was also reported in works such as \cite{ragagnin2019dependency}, \cite{ragagnin2022galaxies}, and \cite{despali2017impact}. 
This observation is further supported by the Median $N^{\rm sub}_{\rm 2D}$ and Median $N^{\rm sub}_{\rm 3D}$ columns in Tables \ref{tab:table1} and \ref{tab:table2} respectively.
Moreover, upon examining \autoref{tab:table2} for \simba, we observe a general decreasing trend in the sub-halo mass (check $M_{\text{sub}}^{2D}$ and $M_{\text{sub}}^{3D}$ from redshift $z=0.394$ to $z=0$).
We have also verified that the same evolutionary trend for the sub-halo mass function is observed when considering a higher redshift snapshot with a value greater than $z=0.394$.
Moreover, the evolution of $R_{\text{200c}}$ for a fixed halo mass between $z=0$ and an arbitrary redshift $z$ is governed by the equation,
$ \frac{R_{\text{200c}}(z=0)}{R_{\text{200c}}(z)} = (H(z)/H_{0})^{2/3}$. This ratio indicates how the volume slightly increases at low redshift as a result of variations in the critical over-density, which can be expressed as
$\Omega_{m}(1+z)^{1/3}+\Omega_{\Lambda}$. This, the pseudo-halo mass evolution, partly leads to the decrease of the satellite number with the $R_{200c}$ at $z=0$. This is because the redshift evolution in \autoref{fig:figure 2} is much larger than this density change.

\section{Cumulative sub-halo $V_{\text{circ}}$ function}\label{Section:Vmax-function}

\begin{figure*}
  \centering
  \includegraphics[width=\textwidth]{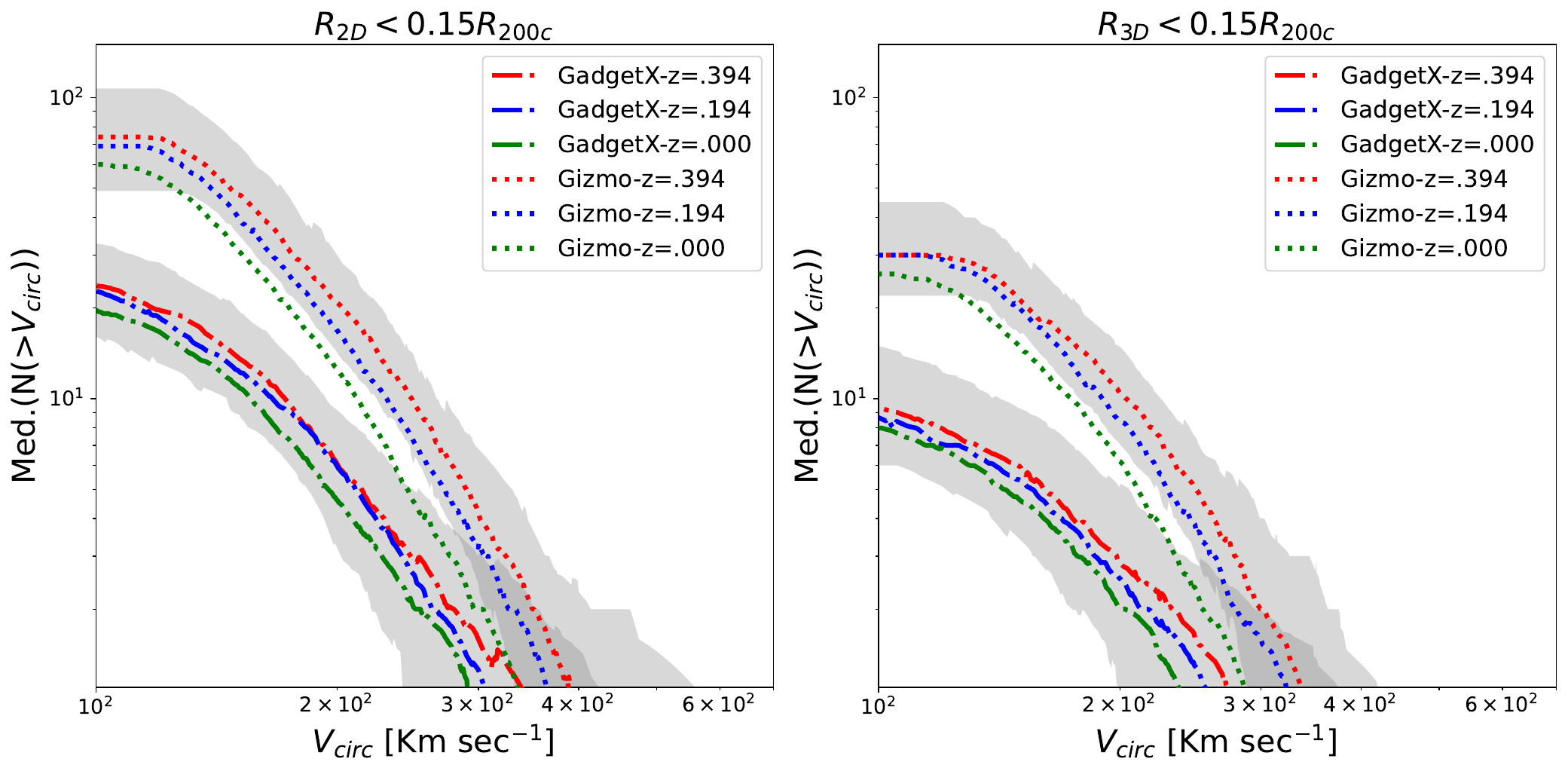}
  \caption{ 2D (left panel) and 3D (right panel) cumulative sub-halo $V_{\text{circ}}$ functions. The dotted line style represents the $\simba$ simulation results, while dash-dot lines show median cumulative sub-halo $V_{\text{circ}}$ functions from \gadgetx. The shaded areas show the $16^{th}-84^{th}$ percentiles from all clusters at $z=0.394$. The $V_{\text{circ}}$ functions of sub-halos in $\gadgetx$ and $\simba$ simulations are displayed for three redshifts: $z=0.394$ (red), $z=0.194$ (blue), and $z=0$ (green). The projected results in the left panel used all sub-halos located within a projected 2D distance of $0.15 R_{\text{200c}}$, i.e., $R_{\text{2D}}<0.15 R_{\text{200c}}$. On the other hand, the right panel illustrates the results using only sub-halos situated within a physical 3D distance of $R_{\text{200c}}$, i.e., $R_{\text{3D}}<0.15 R_{\text{200c}}$.
   We also mention that we employed sub-halos with masses greater than $10^{10} h^{-1} \Msun$ when computing the $V_{\text{circ}}$ functions for both \gadgetx\ and \simba. 
}
  \label{Figure:Vfunction}
\end{figure*}

In this section, we will calculate and compare the cumulative sub-halo $V_{\text{circ}}$ function for both the $\gadgetx$ and $\simba$ simulations at three different redshifts: $z=0.394$, $z=0.194$, and $z=0$. To estimate each sub-halo $V_{\text{circ}}$ in both $\gadgetx$ and $\simba$ simulations, we used the output profiles generated by AHF \citep{knollmann2009ahf}. The output files from the AHF contain radial profiles for halo/sub-halo various properties such as mass, density, rotation curve, escape velocity, etc. Here, we only used the rotation curve of each sub-halo to estimate $V_{\text{circ}}$ in both \gadgetx\ and \simba\ at the three redshifts.
The circular velocity, denoted as $V_{\text{circ}}$ for a sub-halo is determined by identifying the maximum circular velocity at radii greater than zero, ensuring convergence and which is dominated by two-body collisions according to the criterion of \cite{power2003inner}.
The rotation curve for halo/sub-halos is calculated inclusively considering both baryons and dark matter particles in the AHF profile file. We have verified that this value is compatible with the one in the AHF halo properties.

We calculate the sub-halo $V_{\text{circ}}$ function for each host cluster to determine the median cumulative sub-halo $V_{\text{circ}}$ function at the specified redshifts for both $\gadgetx$ and $\simba$ simulations. To calculate the median sub-halo $V_{\text{circ}}$ function, we interpolated the individual sub-halo $V_{\text{circ}}$ functions for each host halo at given $V_{\text{circ}}$ values, and then calculated the median value of $N(>V_{\text{circ}})$ using all the interpolated profiles. This process yields the median cumulative sub-halo $V_{\text{circ}}$ function for the simulated clusters at the respective redshifts.

\autoref{Figure:Vfunction} illustrates the median cumulative sub-halo $V_{\text{circ}}$ function for both $\gadgetx$ and $\simba$ simulations at three different redshifts: $z=0.394$, $z=0.194$, and $z=0$. The left panel shows the function for $R_{\text{2D}}<0.15 R_{\text{200c}}$, while the right panel shows it for $R_{\text{3D}}<0.15 R_{\text{200c}}$.
In \autoref{Figure:Vfunction}, the shaded grey region represents the upper and lower $34\%$ percentiles for clusters at redshift $z=0.394$ in both $\gadgetx$ and $\simba$ simulations. The $V_{\text{circ}}$ functions for $\simba$ are higher compared to $\gadgetx$ for both $R_{\text{2D}}<0.15 R_{\text{200c}}$ and $R_{\text{3D}}<0.15 R_{\text{200c}}$. Once more, we notice that the projection effect leads to an approximately twofold increase in the sub-halo count in both the $\gadgetx$ and $\simba$ simulations. 
Additionally, we noticed a subtle redshift evolution in the cumulative sub-halo $V_{\text{circ}}$ function for both \gadgetx and \simba, which agrees with the result of the sub-halo mass function considering the positive correlation between $V_{\text{circ}}$ and $M_{\text{sub}}$, further discussed below.

\section{\protect$M_{\text{sub}}-V_{\text{circ}}$ relation} 
\label{sec:Section4}
\begin{figure*}
  \centering
  \includegraphics[width=\textwidth]{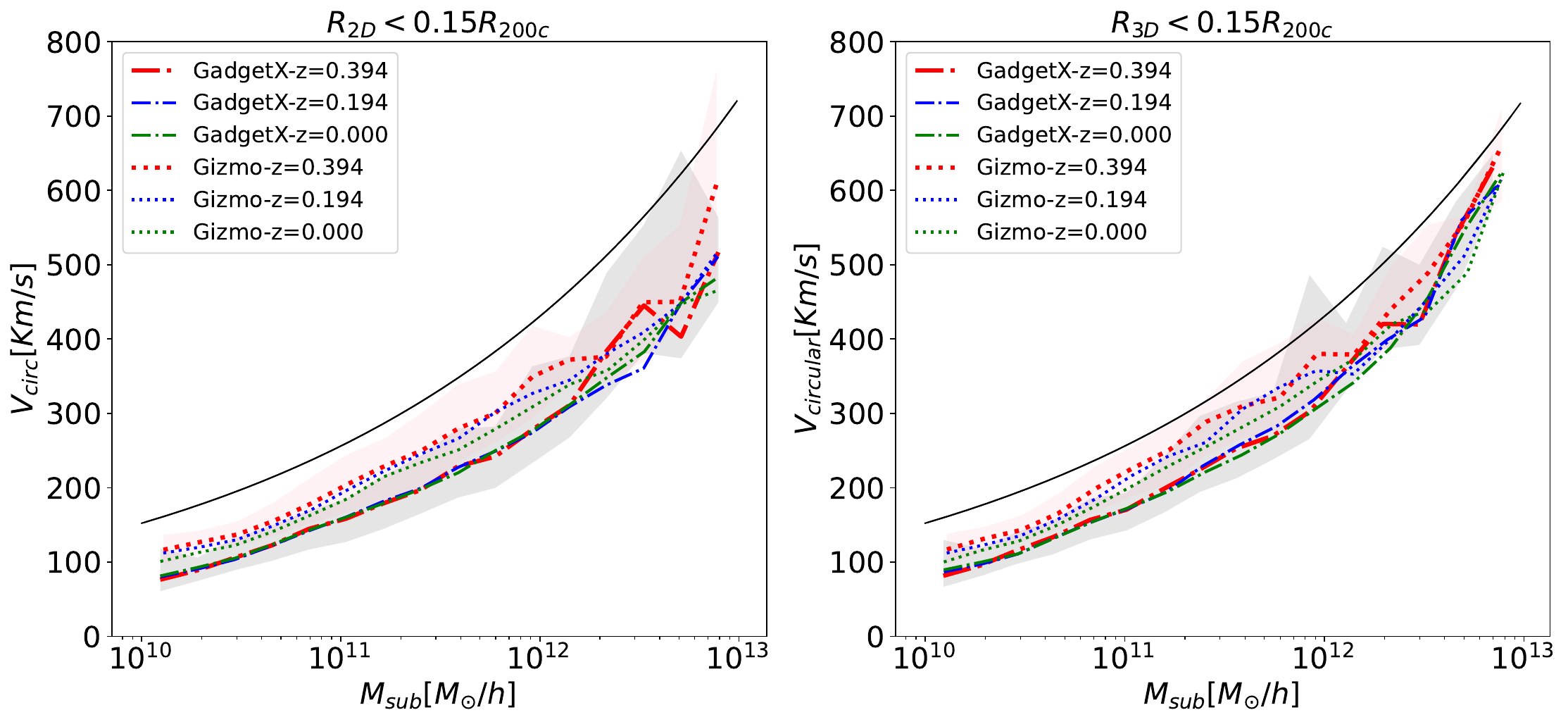}
  \caption{ The relationship between sub-halo mass ($M_{\text{sub}}$) and maximum circular velocity ($V_{\text{circ}}$) for the 2D projected sub-halos on left panel and for the 3D one on the right panel. 
The black solid line is observed fitting relation from \citetalias{meneghetti2020excess} in both panels for reference.
The $M_{\text{sub}}$-$V_{\text{circ}}$ relation for the two simulations is distinguished by distinct line styles, with dashed-dot representing $\gadgetx$ and dotted representing $\simba$. 
The $M_{\text{sub}}$-$V_{\text{circ}}$ relation in $\gadgetx$ and $\simba$ simulations are displayed for three redshifts: $z=0.394$ (red), $z=0.194$ (blue), and $z=0$ (green).
The light grey and red shaded regions depict the upper and lower $34\%$ quantile regions computed in each logarithmic mass bin at redshift $z=0.394$ for $\gadgetx$ and $\simba$, respectively.
}
  \label{Figure:3}
\end{figure*}

In this section, we investigate the discrepancy between the concentration of sub-halos in \theth\ simulations and the lensing results of \citetalias{meneghetti2020excess}.
Following \citetalias{meneghetti2020excess}, \cite{ragagnin2022galaxies, bahe2021strongly}, we employed the $M_{\text{sub}}$-$V_{\text{circ}}$ relationship as a metric to infer the concentration of sub-halos within the clusters. The sample of selected sub-halos for both $\gadgetx$ and $\simba$ remains unchanged.
To derive the $M_{\text{sub}}$-$V_{\text{circ}}$ relationship, we initially divide the sub-halo masses into logarithmic mass bins and subsequently calculate the median $V_{\text{circ}}$ for each respective bin.
$M_{\text{sub}}$-$V_{\text{circ}}$ relation is then obtained by plotting the middle values $M_{\text{sub}}$ in each bin with respect to the corresponding median values of $V_{\text{circ}}$ for each bin. This procedure was repeated for both $\gadgetx$ and $\simba$ simulations at the three redshifts considered in our study.

In \autoref{Figure:3}, we present the $M_{\text{sub}}$-$V_{\text{circ}}$ relationship for the sub-halos from $\gadgetx$ and $\simba$ simulated clusters and compare it with the relation of the observed clusters derived by \citetalias{meneghetti2020excess}. The left panel of \autoref{Figure:3} presents the projected results while the right panel shows the 3D case. We use different colors to show the the $M_{\text{sub}}$-$V_{\text{circ}}$ relations (for both $\gadgetx$ and $\simba$) at three different redshifts, namely $z=0.394, z=0.194$ and $z=0$. 
When sub-halos follow the distance constraint $R_{\text{2D}}<0.15 R_{\text{200c}}$, both $\gadgetx$ and $\simba$ simulated clusters exhibit consistently lower $V_{\text{circ}}$ values compared to the fitting line from observation.
Sub-halos located at the periphery (i.e. $R_{\text{3D}} \approx R_{200c}$) for the $2D$ case $R_{\text{2D}}<R_{\text{200c}}$, cause the simulation's $M_{\text{sub}}$-$V_{\text{circ}}$ relation to shift downward compared to the observed relation. \simba, though, shows slightly higher $V_{\text{circ}}$ than \gadgetx. Furthermore, \simba\ tends to have a weak redshift evolution as a higher $V_{\text{circ}}$ at $z=0.394$ compared to $z=0$, while no redshift evolution is presented in \gadgetx. Similarly, the same conclusions are reached for the case when $R_{\text{3D}} < R_{\text{200c}}$, albeit that both seem to become closer to the observation fitting line, which is in agreement with \citetalias{meneghetti2020excess} and our later correlation studies. Even after considering the sub-halos of the 10 most massive host halos, the discrepancy between the observed and simulated $V_{\text{circ}}$ values persists.
We do not see significant differences between different sub-halo masses regarding the distances to the fitting line, although the shaded regions seem larger, thus closer to the fitting line, at higher sub-halo masses.
We also emphasize that the disparity between the $M_{\text{sub}}$-$V_{\text{circ}}$ relations of \gadgetx\ and \simba\ is primarily limited to the lower sub-halo mass range. For sub-halos with $M_{\text{sub}}>\approx 10^{12} h^{-1} \Msun$, they exhibit a notable degree of agreement.
As we have noticed, the sub-halos at the lower halo mass range can be unresolved.
Based on our analysis of the sampled data, we have not identified any significant deviations or trends in the $M_{\text{sub}}$-$V_{\text{circ}}$ relation that can be directly attributed to unresolved sub-halos in the lower sub-halo mass range. Nevertheless, we acknowledge that further investigations with higher-resolution simulations is necessary to gain a more complete insight into how unresolved sub-halos in this mass range might affect the relation.
We also observed that the simulated $M_{\text{sub}}$-$V_{\text{circ}}$ relationship for sub-halos with masses $M_{\text{sub}}\lesssim10^{11} h^{-1} M_{\odot}$, which is the most crucial mass range for GGSL events \citep{ragagnin2022galaxies}, differs constantly from observations.
Conversely, the $M_{\text{sub}}$-$V_{\text{circ}}$ relation in simulations for the massive sub-halos, $M_{\text{sub}}>4 \times 10^{11} h^{-1} M_{\odot}$, becomes closer to observed relation by varying the baryon parameters, as noted also in \cite{bahe2021strongly}.  However, it's worth noting that this specific range of sub-halo masses is notably higher than what is observed, as highlighted in \cite{ragagnin2022galaxies}.
Though we have a much larger sample and observe that the lines are closer to the observation line at the most massive sub-halo mass range, the discrepancy at the low sub-halo mass end remains unsolved.
Note that the resolution, which could affect this statement for our simulations, given by the checks from \cite{ragagnin2022galaxies}, \cite{bahe2021strongly} and our examinations of the high-resolution the300 clusters, does not significantly impact the $M_{\text{sub}}$-$V_{\text{circ}}$ relation. Therefore, it is still unclear whether this can be solved by varying baryon models or not. The difference between \simba\ and \gadgetx\ suggests that this may be the case.
In the following section, we aim to investigate the influence of sub-halo properties on the $M_{\text{sub}}$-$V_{\text{circ}}$ relation, where we examine how these properties relate to the difference between the $V_{\text{circ}}$ obtained from the simulation and the one derived from observed fitting relations. This difference serves as a measure of the goodness of fit to the $M_{\text{sub}}$-$V_{\text{circ}}$ relation.

\subsection{The effects of the sub-halo properties on $M_{sub}-V_{circ}$ relation}
\begin{figure*}
  \centering
  \includegraphics[width=.95\textwidth]{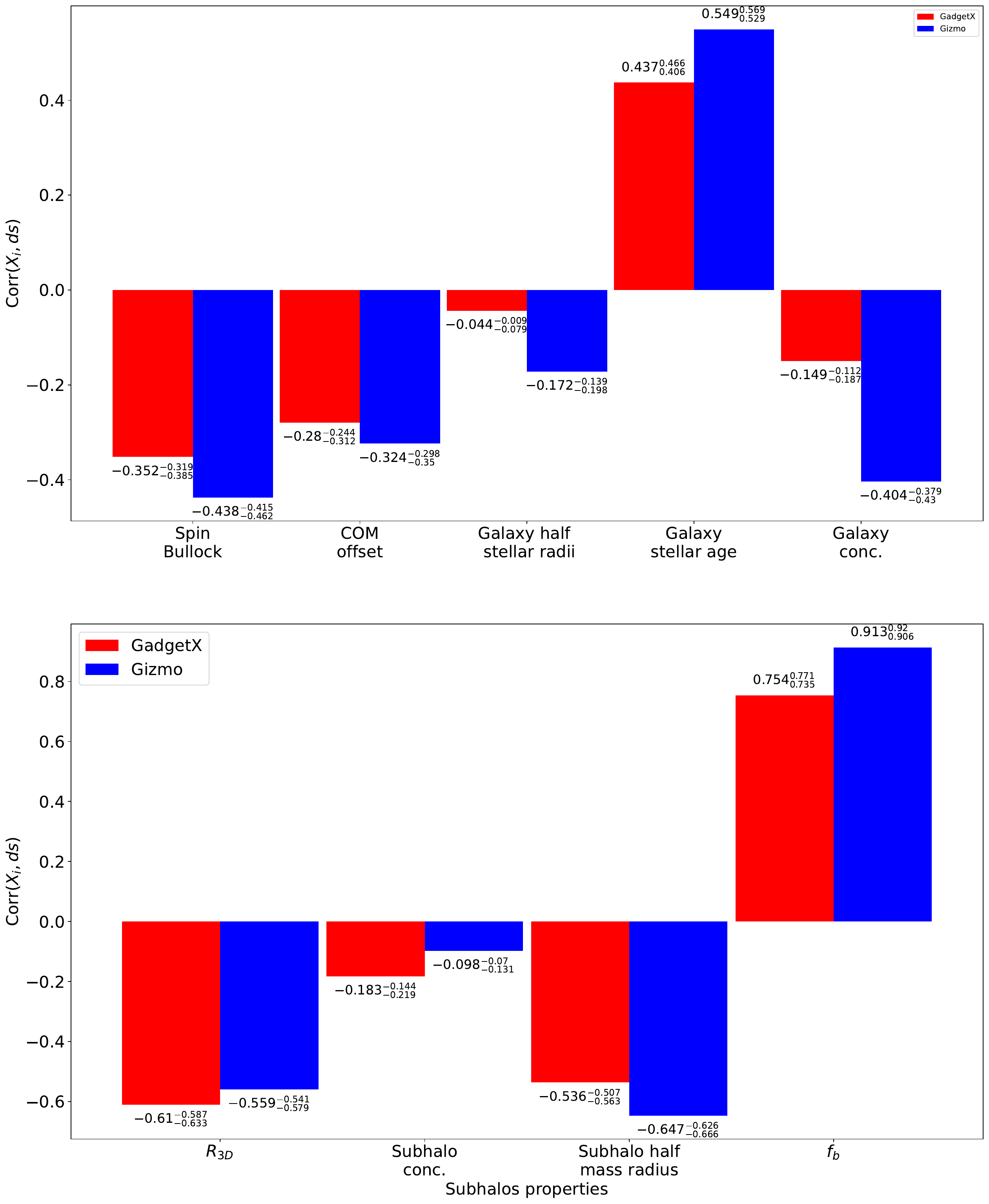}
  \caption{The Spearman correlation coefficient between the physical properties of sub-halos and the residuals $ds$. The residual $ds$ is computed as the distance from the sub-halo's circular velocity obtained from the simulation to the one predicted by the observed relation. To distinguish between the two simulations, we use the red bar plots for the results of $\gadgetx$ and the blue bar plots for the results of $\simba$, respectively. The values of the bar plot are the Spearman correlation coefficient between the sub-halo residual $ds$ and various sub-halo properties.
  Corr($X$,$ds$) defines the Spearman correlation coefficient between the physical property $X$ of sub-halos and the residual $ds$. This parameter is obtained by rank-ordering the sub-halo property $X$ and the residual $ds$, and then calculating the Pearson coefficient based on this rank-order list. The value of this parameter falls between -1 and 1. 
  For the Spearman correlation studies, we chose sub-halos that meet the following criteria: their mass $M_{\text{sub}} > 1.27 \times 10^{11} h^{-1} \Msun$, and their 2D projected distance $R_{\rm 2D}<0.15 R_{200c}$. The Spearman correlation coefficients are accompanied by superscripts and subscripts denoting upper $84\%$ and lower $16\%$ uncertainties, respectively.
  }
  \label{Figure:4}
\end{figure*}

\begin{figure*}
\centering
\begin{subfigure}[b]{\textwidth}
   \includegraphics[width=1\linewidth]{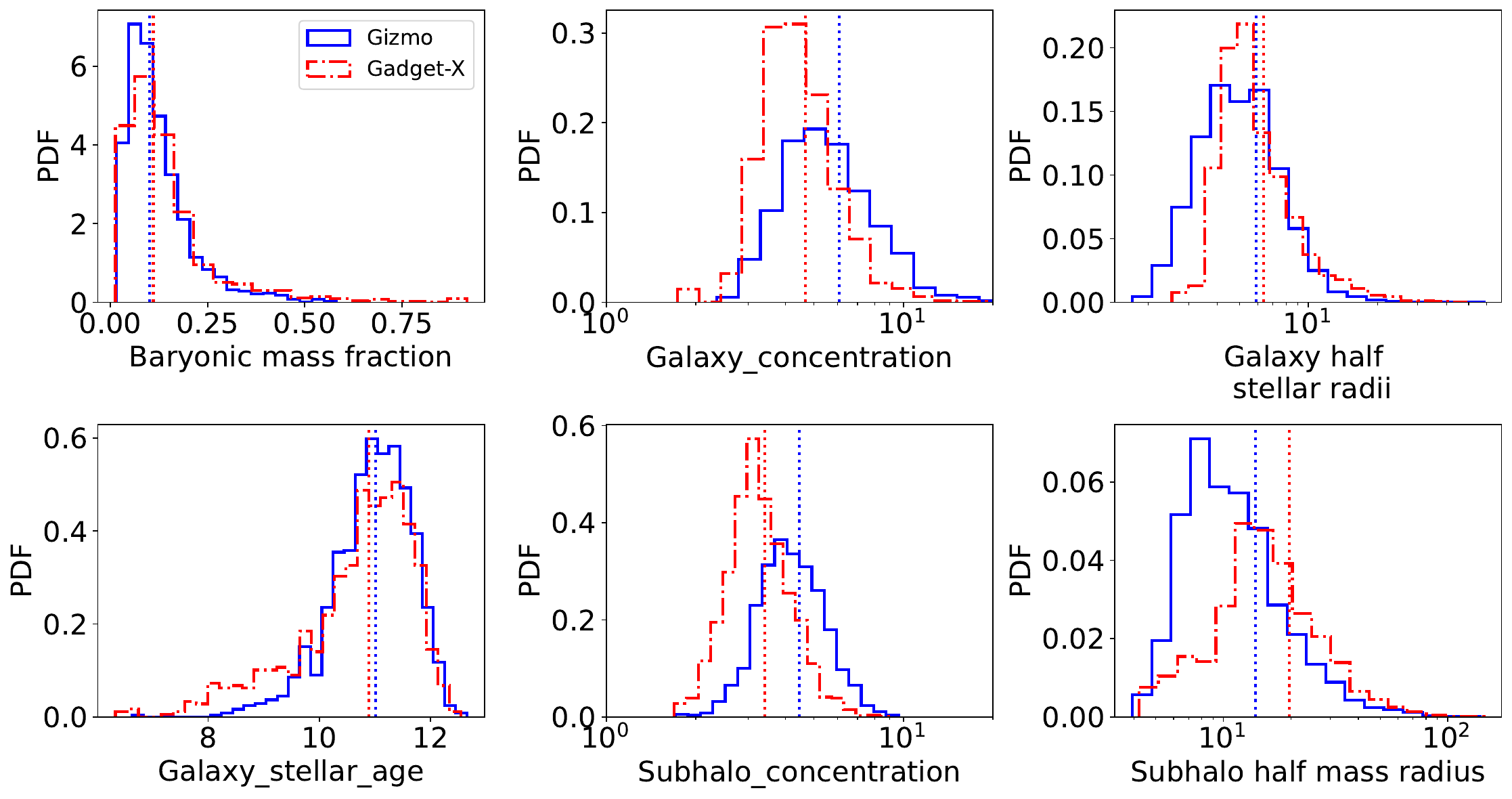}
   
\end{subfigure}

\begin{subfigure}[b]{0.36\textwidth}
   \includegraphics[width=1\linewidth]{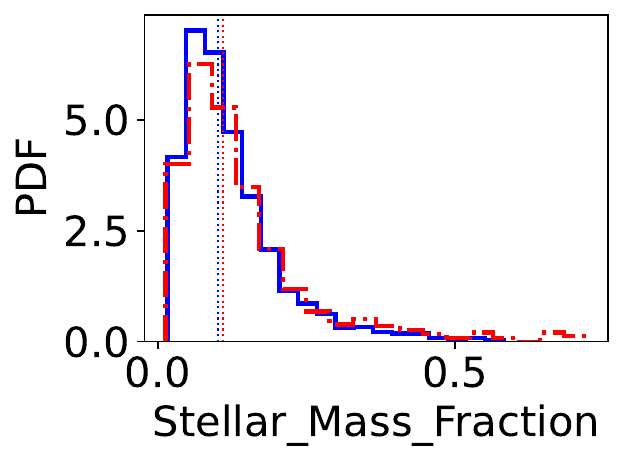}
   
\end{subfigure}
\caption{The probability density functions (PDFs) of the baryonic mass fraction, galaxy concentration, galaxy half stellar radii, galaxy stellar age, sub-halo concentration and sub-halo half mass radii from top left to bottom right for both $\gadgetx$ (red dash-dot steps) and $\simba$ (blue solid steps). The distributions are presented in either linear or logarithms based on their spread ranges. The dotted vertical lines in each plot correspond to the median values of the distributions. For the comparison of sub-halo properties between the two simulations, we selected sub-halos that meet the following criteria: their mass, $M_{\text{sub}} > 1.27 \times 10^{11} h^{-1} \Msun$, and their 2D projected distance, $R_{\rm 2D} < 0.15 R_{200c}$.}
\label{Figure:5}
\end{figure*}

While \simba\ appears to be somewhat closer to the observed fitting line than \gadgetx, the deviation from the observational results remains substantial. This is particularly pronounced in the case of the projected data, which holds greater importance in the observational context. It is interesting to see that different baryon models indeed give slightly different results, which means there may be a cure for this discrepancy by better calibrating the baryon models. Therefore, in order to understand the impact of sub-halo properties on the $M_{\text{sub}}-V_{\text{circ}}$ relation, we perform a Spearman correlation analysis between the different physical properties of the sub-halos and the residual for all sub-halos in the $R_{2D} < 0.15 R_{200c}$ case. The Spearman correlation test involves converting the data into ranks and then calculating the correlation between the ranks of the two variables. 
This Spearman correlation analysis between the different physical properties of the sub-halos and the residual not only provides more statistics, but also presents a consistent comparison to the observation result. The residual $ds$ is calculated for each sub-halo by finding the distance between its $V_{\text{circ}}$ value obtained from the fitting line at its sub-halo mass $V_{\text{circ}}^{\rm fit}$ and the one derived from the simulations $V_{\text{circ}}^{\rm sim}$, it is further normalised to the fitted line value:
\begin{equation}
    ds = \frac{V_{\text{circ}}^{\rm sim} - V_{\text{circ}}^{\rm fit}}{V_{\text{circ}}^{\rm fit}}.
\end{equation}
Note that, we only use sub-halos with $M_{sub}>1.27 \times 10^{11} h^{-1} M_{\odot}$ to calculate these correlation coefficients. 
This is attributed to the potential influence of simulation resolutions on certain sub-halo properties, as sub-halos below this range roughly consist of fewer than 100 dark matter particles.
 Apart from identifying halos and their corresponding sub-halos, AHF \citep{knollmann2009ahf} also provide many physical properties associated with them. Here, we investigate these quantities which should have the most effects on the  $M_{\text{sub}}-V_{\text{circ}}$ relation. 
The sub-halo properties analysed with the Spearman correlation test include the Bullock Spin parameter, which is a measure of the spin of the sub-halo based on \cite{bullock2001profiles}, and the Peebles Spin parameter, which is another measure of the sub-halo's spin based on a different definition by \cite{peebles1969origin}.
The dimensionless spin parameter in \cite{peebles1969origin} is calculated as $\sqrt{E}|J|/G M^{\frac{5}{2}}$, where $E$ represents the total energy, $J$ denotes the angular momentum, and $M$ stands for the mass of the sub-halo or halo. However, estimating this quantity poses challenges as it requires determining the total energy $E$ from simulations and observations. The difficulty arises from the necessity to compute the gravitational potential energy, which, in turn, relies on obtaining accurate information about the mass distribution.
To overcome this problem, an alternative dimensionless spin parameter was proposed in \cite{bullock2001profiles}.
It is calculated as $|J|/\sqrt{2} M R V$, where $|J|$ denotes the angular momentum, $M$ is the mass of the sub-halo, $R$ is the virial radius, and $V$ is the virial circular velocity given by $V = \sqrt{\frac{G M}{R}}$. The measurements of $J, M$, and $V$ are all confined to the virial radius $R$. This makes this spin definition especially attractive since it solely depends on the material within $R$, enabling its calculation for individual components. Hence, using this definition, the radial distribution of the spin is straightforward.

Furthermore, the analysis takes into account the baryonic mass fraction (f\textunderscore b), which represents the proportion of baryonic matter (ordinary matter i.e. gas and stellar content) within the sub-halo. The centre of mass offset parameter (COM\textunderscore offset), the distance between the centre of mass of the sub-halo and its density peak, is also considered. This is normally used as an indicator of the object's dynamical state \citep[see][for example]{Cui2017}.

In addition to that, we further calculate some galaxy and sub-halo properties that may be directly linked to the $ds$, but not provided by AHF. These properties included in the analysis are the physical distance between the host-halo and the sub-halos (R\textunderscore 3D), the galaxy's half-stellar mass radii, the galaxy's stellar age, which is the mass-weighted mean of all-star particles within the half-stellar mass radius, the sub-halo half-mass radii and the galaxy/sub-halo concentrations. As it is very difficult to decide the density profiles for these sub-halos and therefore to estimate their concentration, it is very common to use the ratio of two radii, $R_{80}$ and $R_{20}$, as an indicator of the concentration. Here, $R_{80}$ marks the radius where 80 per cent of the total (stellar) mass of the sub-halo (galaxy) is included. With a similar definition for $R_{20}$,  one would expect a more concentrated density profile should have a higher ratio $R_{80}/R_{20}$.     

The correlation between the physical properties of the sub-halos (for both $\gadgetx$ and $\simba$) and the residual $ds$ is depicted in Fig. \ref{Figure:4}. It is clear that both simulations generally agree on the (anti-)correlation between $ds$ and sub-halo/galaxy properties. Namely, the higher Spin, COM offset, galaxy/sub-halo half mass radius and concentration, further from cluster centre and galaxy concentration, the further distance to the fitted $M_{\text{sub}}-V_{\text{circ}}$ relation. We displayed the Spearman correlation coefficient only for the Bullock spin parameter, which is more robust. The Peebles spin parameter showed a similar trend with a comparable correlation coefficient. At the same time, the older galaxy age (formed earlier) and sub-halos baryonic mass fraction will bring the simulated sub-halo $V_{\text{circ}}$ closer to the observed relation. 
We also examined the correlation trend for the Stellar mass fraction, which exhibited a positive correlation with the residual $ds$. It displayed a closely similar magnitude to the baryon fraction. This is not surprising given that simulated satellites have virtually no gas; therefore, these two fractions are expected to be nearly identical.
It is worth noting that the most significant sub-halo properties are galaxy stellar age, $R_{3D}$ distance, sub-halo half mass radius, and baryon fractions. 
The Spearman correlation trends between $ds$ and $R_{3D}$, as well as $ds$ and $f_b$, obtained from our analysis, have also been reported in \citetalias{meneghetti2020excess} and \cite{bahe2021strongly}, respectively.
The positive correlation between $ds$ and galaxy age suggests early galaxy formation in simulations will provide a better agreement, which is also consistent with the recent JWST observations on the very high-redshift galaxies \citep[see][for example]{Naidu2022,Finkelstein2022}. It is also interesting to note that in the \simba\ simulation, both the gas fraction ($f_b$) and galaxy age are more strongly positively correlated to $ds$ compared to \gadgetx. For the negative correlation between $ds$ and the sub-halo half mass radius, it is very easy to understand: the larger radius, the puffier the sub-halo is, therefore, the lower $V_{\text{circ}}$. Naively, we also expect that the sub-halo half mass radius will be anti-correlated with the sub-halo concentration. By directly looking at the coefficient between the sub-halo half mass radius and concentration, which is also negatively correlated, we suspect that this is caused by different sub-halo masses for the anti-correlation at a fixed sub-halo will be diluted by plotting all the sub-halos together. Therefore, we state that the mentioned correlation between sub-halo half mass radius and concentration is not shown \ref{Figure:5}. Furthermore, this is also applied to the galaxy concentration parameter. This simple definition of concentration may not serve our purpose well here. We also emphasize that this correlation study holds greater relevance in the context of sub-halos with $M_{\text{sub}} > 1.27 \times 10^{11} h^{-1} \Msun$, as we applied this mass threshold to mitigate resolution-related effects that can influence the physical properties of sub-halos. Hence, no definitive conclusions can be drawn regarding the influence of sub-halo properties on the $M_{\text{sub}}-V_{\text{circ}}$ relation for sub-halos falling below the mentioned mass threshold.

\bigskip

In addition to examining the correlations with $ds$, which highlights the individual sub-halo properties' effect, we also compared the distribution of sub-halo properties between $\gadgetx$ and $\simba$ in \autoref{Figure:5}. The distributions of galaxy/sub-halo properties are presented with the 1D probability density functions for both simulations. Through these comparisons, we expected to further understand the model differences between the two simulations and how they impact the $M_{\text{sub}}-V_{\text{circ}}$ relation. In \autoref{Figure:5}, only 6 interesting and important sub-halo properties are picked to show. 

First, the sub-halos in the simulated clusters of $\gadgetx$ contain a marginally higher amount of baryonic content compared to those in $\simba$. Note that the distribution difference is larger when including low-mass sub-halos. The positive correlation illustrated in \autoref{Figure:4} indicates that as the baryon fraction increases, the simulation's $M_{\text{sub}}$-$V_{\text{circ}}$ relation aligns more closely with the observed fitting relation. The explanation behind this is: the inclusion of baryons through tidal stripping leads to an observed offset towards higher $V_{\text{circ}}$ values in the $M_{\text{sub}}-V_{\text{circ}}$ relation \citep{bahe2021strongly}, which is also supported by the presence of sub-haloes with increased baryonic content results from the removal of dark matter in galaxies with the stellar mass is largely preserved (\cite{armitage2019cluster,bahe2019disruption, joshi2019trajectories}). However, this result seems to contradict the conclusion that \simba\ is closer to the fitting line than \gadgetx\, while their baryon fractions are very similar. We suspect the baryon fraction is only a sufficient condition, not a necessary condition to bring up the $V_{\text{circ}}$. Similar to the baryon fraction, the galaxy age distributions between \gadgetx\ and \simba\ are also very similar with a slight excess of young galaxies in \gadgetx. Therefore, the similarity of the two sub-halo properties between \gadgetx\ and \simba\ indicates that other quantity differences are the key to explaining the differences in the $M_{\text{sub}}-V_{\text{circ}}$ relation. They are the sub-halo/galaxy half-mass radii and concentrations: it is clear that \simba\ has smaller half-mass radii, thus a higher concentration of both galaxy and sub-halo compared to \gadgetx. This is in agreement with Meneghetti et al. in prep. which found that the GGLS signal is also higher in \simba\ than \gadgetx, albeit that is still about a few times lower than observation. To boost the $V_{\text{circ}}$, as well as the GGLS signal, we will need even more compact sub-halos/galaxies. To achieve that goal, we suspect an even earlier galaxy formation may bring the simulation closer to observation.

\subsection{Global cluster properties impact on $M_{sub}-V_{circ}$ relation}

\begin{figure*}
  \centering
  \includegraphics[width=\textwidth]{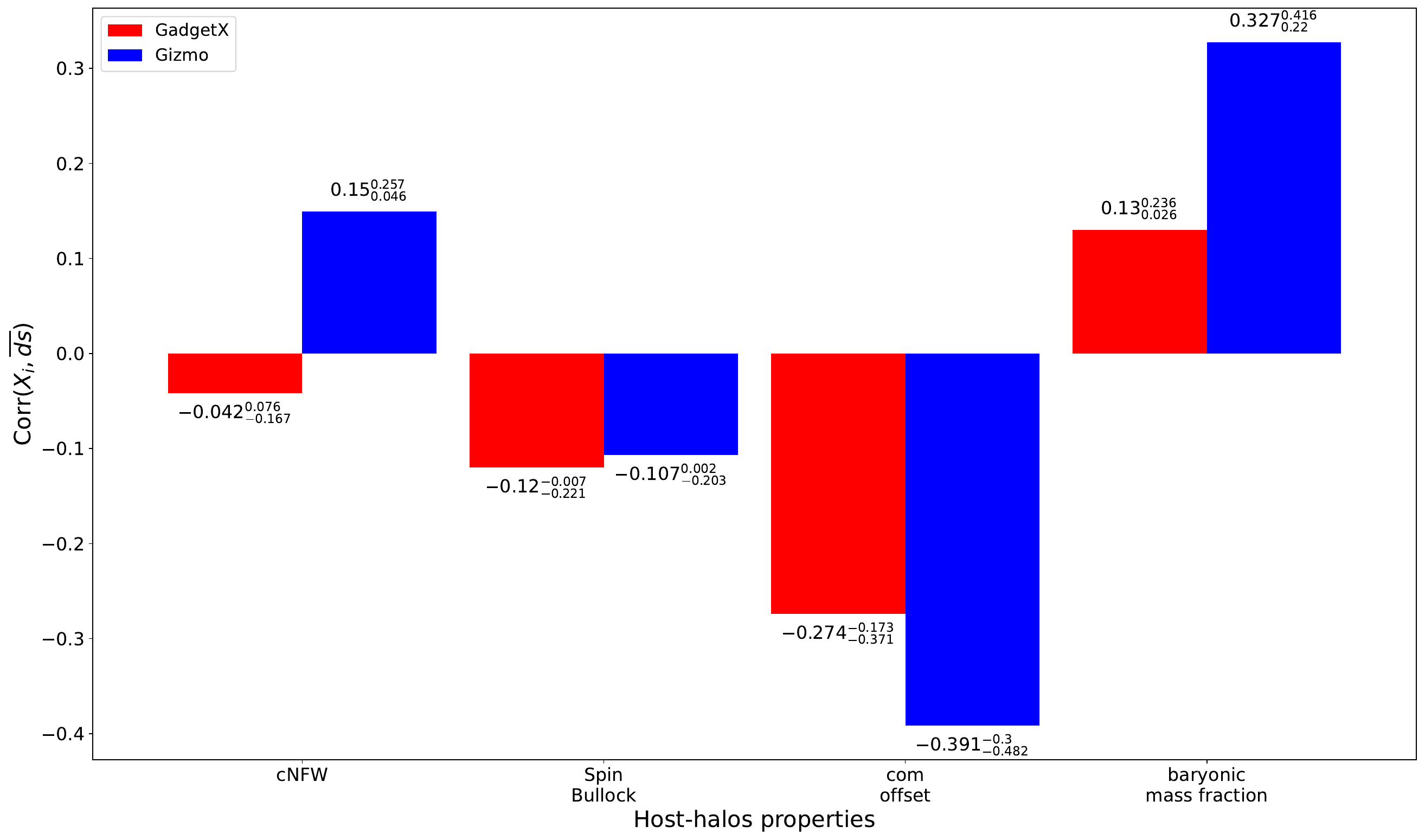}
  \caption{Similar to \autoref{Figure:4}, but for the correlation between the cluster properties and global residuals ($\overline{ds}$). This correlation can be used to infer the impact of cluster/host-halo physical properties on the $M_{\text{sub}}-V_{\text{circ}}$ relationship. Once again, we selected sub-halos from the host clusters that met the following criteria: their mass, $M_{\text{sub}} > 1.27 \times 10^{11} h^{-1} \Msun$, and their 2D projected distance, $R_{\rm 2D} < 0.15 R_{200c}$.
 }
  \label{Figure:6}
\end{figure*}

The next step in our analysis is to investigate the influence of the global properties of the host halo on the $M_{\text{sub}}-V_{\text{circ}}$ relationship. This investigation is to provide some hints on whether the selected clusters in observation are biased or not. In order to determine any potential impact, we provide a similar study on the Spearman coefficient between the physical properties of the host halos and the global residual $\overline{ds}$. Here, the global residual $\overline{ds}$ for each host halo is computed by averaging all its sub-halos' $ds$, which are measured in the previous section. 

In \autoref{Figure:6}, we show the coefficient between $\overline{ds}$ and these four selected cluster properties: cNFW, Bullock spin parameter, COM offset and total baryon fractions. 
Additionally, the analysis considers the Navarro-Frenk-White profile \citep{navarro1996structure} dimensionless concentration parameter (cNFW), which characterises the concentration of the sub-halo's density profile.
The concentration parameter, denoted as cNFW, is typically determined by fitting a Navarro-Frenk-White profile to the halo density. It describes how the density of the halo changes with radial distance from its centre. Here, we simply use the concentration parameter in AHF calculated by following the approach of \cite{prada2012halo}. They utilise the circular velocity ($V_{\text{circ}}$) and the circular velocity at the virial radius, which is defined in terms of the halo's virial mass and radii. All the other halo properties are introduced in the previous section.

Besides the cNFW parameter, the two simulations show similar correlations with the $\overline{ds}$. \gadgetx\ suggests that concentrated halos tend to give a lower $M_{\text{sub}}-V_{\text{circ}}$ relation, while \simba\ suggests the opposite. However, neither shows a strong relation with $\overline{ds}$. Both Bullock and Peebles defined spin parameters negatively correlate with $\overline{ds}$ indicating slow-rotating halos tend to be closer to the observed fitting line. We report the more robust Bullock spin parameter in \autoref{Figure:6}.
 Again, the correlation is not very strong. The highest coefficient is the COM, which suggests that the relaxed halos tend to agree with observation better. This can be understood as this: the relaxed cluster tends to form earlier \citep[see][for example, for the relations of cluster dynamical state with halo formation time and concentration]{Mostoghiu2019}, and the sub-halos inside tend to have a longer time for stripping, thus only the core regions are remaining, which will have a shorter half-mass radius with higher $V_{\text{circ}}$. However, it is worth note that one cluster, MACSJ0416, in \citetalias{meneghetti2020excess}, seems to be unrelaxed. 
This seems contract to our previous prediction. However, we argue that the majority of the sample in \citetalias{meneghetti2020excess} (see also \cite{Meneghetti2022}) also are more relaxed. While the simulation sample is more balanced, see \cite{deluca2021,Zhang2022}.

The positive correlation between $\overline{ds}$ and the halo baryon fraction is in agreement with the correlation result for sub-halos. It is expected that the higher halo baryon fraction connects with a higher sub-halo baryon fraction. However, it is unclear which is the determined reason: the baryon-rich halo merged into the host halo to bring more baryons or the host halo is baryon rich with the sub-halos can retain their baryon longer. It is naturally to think that a higher halo baryon fraction would induce stronger ram pressure with potentially stronger tidal forces, thus leads to a lower baryon fraction in the sub-haloes. It is known recently that the gas in the infalling halos is easily stripped out \citep{Haggar2020}, even before reaching the virial radius of the cluster, this can also happen to the infalling groups as well \citep{Haggar2023}. There the baryon fraction for the satellite galaxies are dominated by stars. On the other hand, the galaxy is more concentrated compared to dark matter, thus less easy get stripped (see Contreras-Santos 2023 in prep.). Therefore, the two high baryon fractions are actually consistent, because the stronger stripping will only remove more dark matter particles and result in a higher subhalo baryon fraction. 

\section{Conclusions} \label{sec:Section5}
 The study by \citetalias{meneghetti2020excess} examined the gravitational lensing properties of galaxy clusters and their sub-halos, revealing a significant discrepancy between observed clusters and hydrodynamic simulations within the $\Lambda$CDM cosmology. Notably, observed clusters exhibited a much higher probability of Galaxy-Galaxy Strong Lensing (GGSL) than simulated clusters.
 Moreover, they utilized maximum circular velocities ($V_{\text{circ}}$) of sub-halos as a metric to assess compactness, finding that sub-halos in observed clusters
 had higher $V_{\text{circ}}$ values compared to those in mass-matched clusters from simulations. This suggests that galaxies in observed clusters are more efficient at lensing background sources and are more compact than those in the simulations. In this study, we thoroughly investigated the discrepancy between the simulations and observations discussed in\citetalias{meneghetti2020excess}.

In our study, we used simulated clusters from the $\theth$ project \cite{cui2018three,cui2022three} with masses $M_{200} > 6.5 \times 10^{14} h^{-1} M_{\odot}$. We aimed to compare these simulated clusters with the observations of three primary reference clusters of \citetalias{meneghetti2020excess} that have a median redshift of $z=0.39$.
We selected a sample of 90 host clusters from the $\gadgetx$ simulation and 82 host clusters from the $\simba$ simulation at a redshift of $z=0.394$ to compare it fairly with observations of \citetalias{meneghetti2020excess}.
We then expanded our analysis by including host clusters at two additional redshifts: $z=0.194$ and $z=0$ for evolutionary studies. The selected clusters at $z=0.194$ for $\gadgetx$ and $\simba$ are 180 and 169, respectively. Similarly, at redshift $z=0$, $\gadgetx$ and $\simba$ provide 321 and 302 cluster samples. Further details about our sample of selected clusters and their sub-halos can be found in Table \ref{tab:table1} and Table \ref{tab:table2}. 
In our analysis, we found the following:
\begin{itemize}
\item The cumulative sub-halo mass function shows an overall consistency between MACSJ0416 and MACSJ1206 clusters from \citetalias{meneghetti2020excess} and the $\simba$ simulation with $R_{\text{2D}} < 0.15 R_{\text{200c}}$  (\autoref{Figure:1}, left). However, for $\gadgetx$, agreement between the observation and simulation is only found at a higher sub-halos mass range. The discrepancy at the low-mass end is attributed to a stronger resolution dependence in the baryon model of $\gadgetx$. The $2D$ vs $3D$ comparison of the sub-halo mass function (as shown in \autoref{Figure:1}) highlights the substantial impact of projection effects, revealing a two-fold increase in sub-halo numbers in 2D compared to 3D.
\item The redshift evolution study of cumulative sub-halos mass function reveals that while the median halo mass increases with decreasing redshift, the number of sub-halos within massive clusters decreases toward the present time. 
The analysis of the normalised sub-halo mass function shows a clear redshift evolution, where a greater number of sub-halos are expected to be observed at earlier times when they are less concentrated within their host halos. The sub-halo mass function for both $\gadgetx$ and $\simba$ at $z=0$ is lower (fewer sub-halos) compared to $z=0.394$, indicating a decrease in the number of sub-halos within host halos over time \autoref{fig:figure 2}.
\item Both \gadgetx\ and \simba\ simulations consistently show lower circular $V_{\text{circ}}$ for sub-halos compared to the fitting line obtained from observations when following the distance constraint $R_{\text{2D}} < 0.15 R_{\text{200c}}$. However, \simba\ exhibits slightly higher $V_{\text{circ}}$ values than \gadgetx. Furthermore, \simba\ shows a weak redshift evolution with higher $V_{\text{circ}}$ at $z=0.394$ compared to $z=0$, unlike $\gadgetx$.
\item The $M_{\text{sub}}$-$V_{\text{circ}}$ relationship for sub-halos with masses $M_{\text{sub}} < 10^{11} h^{-1} M_{\odot}$ shows a noticeable difference between observations and simulations. This discrepancy is particularly relevant in the context of GGSL.
On the other hand, when considering massive sub-halos with $M_{\text{sub}} > 4 \times 10^{11} h^{-1} M_{\odot}$, where simulations are a little closer to observed fitting relation, albeit not in perfect agreement, the significance of the discrepancy decreases due to the limited number of observed sub-halos within this mass range. Therefore, in this range of sub-halo masses, the observed fitting relation of \citetalias{meneghetti2020excess} is extrapolated.
The contrasting results obtained from the $\gadgetx$ and $\simba$ simulations indicate potential to address this issue by fine-tuning the baryon models used in the simulations. However, as shown by \cite{Meneghetti2022}, this fine-tuning is difficult to achieve on the mass scales relevant for GGSL without creating inconsistencies with observations at higher masses. For example, simulations with high star formation efficiency and/or lower energy feedback from AGNs produce an excess of galaxies with masses $\gtrsim 10^{12}\; M_\odot$ compared to observations. Meneghetti et al. (in prep.) noted that the \simba simulations exhibit this problem.
\item The Spearman correlation analysis between sub-halo/galaxy properties and the residual $ds$ reveals that both simulations agree that there is a correlation or anti-correlation between $ds$ and various sub-halo/galaxy properties \autoref{Figure:4}. The significant sub-halo properties that notably impact the residual $ds$ are the galaxy stellar age, distance from the cluster's centre ($R_{3D}$), sub-halo half mass radius, and baryon fraction. The Spearman correlation value indicates that the sub-halo half mass radius and being further away from the cluster centre is associated with a more significant deviation from the observed $M_{\text{sub}}-V_{\text{circ}}$ relation. On the other hand, older galaxy stellar age (formed earlier) and higher sub-halo baryonic mass fraction tend to bring the simulated sub-halo $V_{\text{circ}}$ closer to the observed relation.
\item Upon comparing the sub-halo properties of $\gadgetx$ and $\simba$, it is evident that $\gadgetx$ exhibits slightly higher baryonic content in its simulated clusters' sub-halos (\autoref{Figure:5}). Additionally, the distribution of galaxy ages is highly comparable between \gadgetx\ and \simba, with a slightly higher proportion of young galaxies in $\gadgetx$ (\autoref{Figure:5}).
From the Spearman correlation analysis of sub-halo properties, we would anticipate that \gadgetx\ will be closer to the observation fitting line compared to \simba; however, we observed the opposite.
Specifically, the size and concentration of sub-halos/galaxies are identified as crucial factors that contribute to the differences in the $M_{\text{sub}}-V_{\text{circ}}$ relation, with $\simba$ having smaller sizes and higher concentrations compared to $\gadgetx$. The differences in sub-halo properties imply that creating even more compact sub-halos/galaxies, maybe through earlier galaxy formation, may result in improved model-observational data alignment.
\item Investigation of global host halo properties in relation to the $M_{\text{sub}}$-$V_{\text{circ}}$ relationship reveals that relaxed halos exhibit the strongest alignment with observations (negative correlation with COM offset and $\overline{ds}$). A modest negative correlation between spin parameters and $\overline{ds}$ indicates a tendency for slow-rotating halos to be closer to the observed fitting line, albeit with a weak correlation. Additionally, positive correlation is observed between $\overline{ds}$ and the halo baryon fraction, suggesting a connection to the baryon fraction of sub-halos \autoref{Figure:6}.
\end{itemize}

In conclusion, our analysis of galaxy clusters simulated using both $\gadgetx$ and $\simba$ in the $\theth$ project reveals a discrepancy when comparing them to observations of \citetalias{meneghetti2020excess}. Our findings suggest that some contemporary simulations struggle to faithfully replicate the observed abundance and compactness of sub-halos. This disparity may arise from limitations in baryonic modeling, systematic challenges within our simulation approaches, uncertainties in observational data and their modeling, or potentially, limitations inherent to the $\Lambda$CDM framework.

It is necessary to note here that the comparison done in this paper is based on the AHF halo catalogue instead of SUBFIND in previous studies. We refer to \cite{Onions2012} and \cite{castro2023euclid} for detailed comparisons between different sub-halo finders and discussions. Both AHF and SUBFIND have unbinding processes to remove the particles that are not gravitationally bound to the sub-halo. This is inconsistent with the sub-halo mass measured in observation apart from the projection effect. Nevertheless, using the observation-like sub-halo mass, will only increase the discrepancy between simulation and observation for both will increase the sub-halo mass and shift the simulated $M_{\text{sub}}$-$V_{\text{circ}}$ towards the right side, i.e. away from the observed fitting line. 

\section*{Acknowledgements}
The authors would like to express their sincere gratitude to Frazer Pearce and Elena Rasia for the insightful discussion and helpful comments, which significantly improved the analytical aspect of this work. WC is supported by the STFC AGP Grant ST/V000594/1 and the Atracci\'{o}n de Talento Contract no. 2020-T1/TIC-19882 granted by the Comunidad de Madrid in Spain. He also thanks the Ministerio de Ciencia e Innovación (Spain) for financial support under Project grant PID2021-122603NB-C21 and ERC: HORIZON-TMA-MSCA-SE for supporting the LACEGAL-III project with grant number 101086388. Carlo Giocoli thanks the support from INAF theory Grant 2022: Illuminating Dark Matter using Weak Lensing by Cluster Satellites. 

The high-resolution simulations were performed at the MareNostrum Supercomputer of the BSC-CNS through The Red Española de Supercomputación grants (AECT-2022-3-0027, AECT-2023-1-0013), and at the DIaL -- DiRAC machines at the University of Leicester through the RAC15 grant: Seedcorn/ACTP317

\section*{Data Availability}

The data underlying this article will be shared on reasonable request to the corresponding author. The simulation data is provided by the300 collaboration which can also be accessed upon request on their website.
 



\bibliographystyle{mnras}
\bibliography{reference} 








\bsp	
\label{lastpage}
\end{document}